\documentclass[paper]{JHEP}
\usepackage[centertags]{amsmath}
\usepackage{amsfonts} \usepackage{amssymb} \usepackage{amsthm}
\usepackage{graphicx}
\usepackage{psfrag}

\newcommand{\ket}[1]{|{#1}\rangle}
\def\one{{\rm 1\kern -.9mm l}}

\def\beq{\begin{equation}}
\def\eeq{\end{equation}}
\def\beqa{\begin{eqnarray}}
\def\eeqa{\end{eqnarray}}

\newcommand{\comm}[2]{\left[#1,#2\right]}
\newcommand{\Tr}{\mathrm{Tr}\,}
\newcommand{\uno}{\mbox{1\!\negmedspace1}}
\def\rr{(\rho^2+|x|^2)}
\def\tr{{\rm tr}\,}

\def\NS{{\rm NS}}
\def\R{{\rm R}}

\def\ppz{{(0)}}

\newcommand{\lvev}{\langle\hskip -6pt\langle\hskip 4pt}
\newcommand{\rvev}{\hskip 4pt\rangle\hskip -6pt\rangle}
\newcommand{\bea}{\begin{eqnarray}}
\newcommand{\ena}{\end{eqnarray}}

\usepackage{graphicx}
\usepackage{dcolumn}
\def\ii{\mathrm{i}}
\def\ee{\mathrm{e}}
\title{Non-commutative (D)-instantons~\thanks{Work partially supported
by the European Community's Human Potential Programme under contract MRTN-CT-2004-005104
``{Constituents, Fundamental Forces and Symmetries
of the Universe}'', and by the Italian M.I.U.R under contract
PRIN-2003023852 ``{Physics of fundamental interactions: gauge theories,
gravity and strings}''.}}
\author{
Marco Bill\'o
,
Marialuisa Frau
,
Stefano Sciuto, Giuseppe Vallone
\\
Dipartimento di Fisica Teorica, Universit\`a di Torino\\
and Istituto Nazionale di Fisica Nucleare - sezione di Torino \\
via P. Giuria 1, I-10125 Torino, Italy
}
\author{
Alberto Lerda%
\thanks{E-mails: \tt{billo,frau,sciuto,vallone,lerda@to.infn.it}}\\
Dipartimento di Scienze e Tecnologie Avanzate \\
Universit\`a del Piemonte Orientale, I-15100 Alessandria, Italy\\
and Istituto Nazionale di Fisica Nucleare  - sezione di Torino \\
via P. Giuria 1, I-10125 Torino, Italy
}

\abstract{We study systems of D3 and D$(-1)$ branes in a NS-NS magnetic background
and show that, when the brane configuration is stable, the physical degrees of freedom of the open
strings with at least one end-point on the D-instantons describe the ADHM moduli of instantons
for non-commutative gauge theories. We also prove that disk diagrams with
mixed boundary conditions are the sources for the classical profile of the
non-commutative gauge instantons in the singular gauge.
We finally compare the string theory description in a large distance expansion
with the non-commutative ADHM construction in the singular gauge
and find complete agreement at perturbative level in the non-commutativity parameter.}

\preprint{DFTT-35/2005}

\begin{document}
\section{Introduction}
\label{sec:intro}
Quantum field theories on non-commutative spaces display a very
rich spectrum of unusual properties and for this reason they have attracted a wide
interest in the last few years. For instance, they contain a minimal
distance scale $|\theta|$, provided by the
non-commutativity parameter $\theta^{\mu\nu}\sim\comm{x^\mu}{x^\nu}$, which naturally cuts
off the theory in the UV,
though often at the price of peculiar UV/IR mixing effects.

Initially, a lot of work was devoted to the analysis of non-commutative theories in a
field-theoretic framework, but an even
greater attention was sparkled by the realization
that non-commutativity arises most naturally in a
string theory set-up. The stringy connection was originally pointed
out \cite{Connes:1997cr} in the context of (M)atrix theory compactification, but
it was subsequently\,%
\footnote{Much ground-work having already been performed,
quite earlier, in Ref. \cite{Abouelsaood:1986gd}.}
established in a more direct way
\cite{Sheikh-Jabbari:1997yi}-\cite{Chu:2000pc} by considering
open strings in a magnetic field
or in a closed string background with a non-trivial flux for the  $B_{\mu\nu}$ field of the NS-NS
sector\,%
\footnote{Constant fluxes of RR fields lead instead to
non anti-commutative theories
\cite{deBoer:2003dn}-\cite{Seiberg:2003yz}.}.

A very interesting aspect of the non-commutative deformations of gauge theories is the study of
their effects on instantons. This is the main subject of this paper.
Realizing a four-dimensional $\mathrm{U}(N)$ gauge theory by a stack of $N$
coincident D3 branes, instantons with charge $k$ can be obtained by
adding $k$ D$(-1)$ branes, also known as D-instantons
\cite{Witten:1995im,Douglas,Seiberg:1999vs}. This brane
realization not only reproduces and physically explains the ADHM \cite{Atiyah:ri}
construction (see, for instance, Ref.
\cite{Dorey:2002ik} and references therein); it also
accounts for the profile of the classical solution and for the instanton calculus
of correlators from a string theory point of view, as shown in Ref.
\cite{Billo:2002hm}, building also on techniques already introduced in Ref. \cite{Green:2000ke}.
In a trivial background the moduli space of these stringy instantons coincides with that of
classical ADHM gauge instantons \cite{Seiberg:1999vs}, but when we introduce a
non-commutative deformation $\theta_{\mu\nu}$ by turning on a $B_{\mu\nu}$ field,
it is deformed and no longer coincides with the classical one unless
the background is self-dual
(or anti-self-dual in the case of anti-instantons) \cite{Seiberg:1999vs}.
In particular, the ADHM constraints
are modified in such a way that the small-instanton singularity
\cite{Witten:1995gx}, which corresponds to the possibility of detaching the D-instantons
from the world-volume of the D3 branes, is removed.
This basically happens because the D3/D$(-1)$ system no longer satisfy a
no-force condition and is no longer stable when the background field is not
(anti-)self-dual, in perfect agreement
with the features of the (anti-)instanton solutions on non-commutative $\mathbb{R}^4$
obtained by extending the usual ADHM construction to the non-commutative
gauge theories \cite{Nekrasov:1998ss}-\cite{Wimmer:2005bz}.

In this paper, we intend to pursue the line of thought of Ref. \cite{Billo:2002hm},
already successfully applied
in the case of non anti-commutative deformations \cite{Billo:2004zq}, and explicitly derive the
measure on moduli space
and the instanton profile from open string disk amplitudes in presence of the
NS-NS $B$ background. In doing so, we retrieve the expected behaviour, but along the way
we encounter some crucial subtleties that render the entire construction non-trivial.

After presenting in section \ref{sec:d3d-1} a brief review
of the stringy ADHM construction, in section \ref{sec:d3d-1wb} we analyze the quantization of the open
strings of the D3/D$(-1)$ system in a $B$ background.
As is well known, this quantization can be carried out exactly in the RNS
formalism\,%
\footnote{This is in contrast with the non anti-commutative case where the RR background
can only be inserted perturbatively,
even if, in the end, this turns out to be sufficient, see for
instance \cite{Billo:2004zq,Billo:2005jw}.}
for all kinds of open strings (namely those stretching between two D3's, or between two D$(-1)$'s,
or the mixed ones). However, a careful analysis reveals that
there exist different possibilities of imposing boundary
conditions on the world-sheet fermions that are compatible with
the $B$ background.
One therefore obtains a number of different open string sectors that is larger than naively
expected. Just like in the commutative case, also in the presence
of $B$ the physical excitations of open strings with at least one
end-point on the D$(-1)$ branes are interpreted as the instanton ADHM moduli; however
in the non-commutative case two points should be stressed: first,
this identification is possible only for a self-dual background (or for an anti-self-dual background in the case
of anti-instantons), {\it i.e.} when the D3/D$(-1)$ system is stable;
second, the precise relationship between open string states
and ADHM moduli is highly non-trivial and actually it requires a suitable
combination of the various open string sectors associated with the different types of
fermionic boundary conditions mentioned above.

In section \ref{secn:solution}, we consider a self-dual background
and, using the string construction of section \ref{sec:d3d-1wb}, prove that, as expected, the moduli space for non-commutative
instantons is not modified with respect to the commutative one. Moreover, we show that the leading
term in the large distance expansion of the classical
solution is generated by the gauge boson emission amplitude from mixed disks, in perfect analogy
with the ordinary gauge instantons \cite{Billo:2002hm}.
The results of these string calculations are then compared with the
non-commutative field theory expectations in section \ref{sec:nci_moyal}.
Since the classical instanton profile obtained from the mixed disk emission
diagram is in the so-called ``singular
gauge'', in order to compare it with the non-commutative ADHM construction
we have to describe the latter also in the singular gauge, rather than in
the regular one as is usually done in the literature. This is not a trivial task (see,
however,
Ref. \cite{Tian:2002si} for a discussion on this point),
but in our context it is enough to write down an instanton
profile which solves the ADHM constraints
up to some exponentially suppressed contributions, and
allows a meaningful and successful comparison with the string results of
section \ref{secn:solution}.

We conclude by considering the D3/D$(-1)$ system in a generic
({\it i.e.} non (anti-)
self-dual) $B$ background, whose effects
can only be treated in a perturbative way.
For the 3/3 strings, this approach was in fact exploited already
in \cite{Schomerus:1999ug} to show
the emerging of a non-commutative gauge theory from string theory amplitudes.
For the $(-1)$/3 and the $(-1)$/$(-1)$ strings things are actually simpler
and the computation of mixed open/closed string diagrams with a single $B$-insertion is sufficient
to exhibit the expected deformation of classical instanton moduli space and of
the ADHM constraints.
Finally, in the appendix we list
our notations and conventions.

\vskip 0.8cm
\section{Gauge instantons from D3/D(--1) systems}
\label{sec:d3d-1}

The ADHM construction \cite{Atiyah:ri} of supersymmetric gauge instantons and
their moduli space can be derived in full
detail from string theory by considering systems of D3 branes
and D-instantons (for a review see, for instance,
\cite{Dorey:2002ik})~\footnote{For anti-instantons one should consider instead
anti-D($-1$) branes.}. In this approach, the
auxiliary variables of the ADHM construction correspond to the degrees
of freedom associated to open strings with at least one end-point attached to
the D-instantons, and the measure on the moduli space as well as
the instanton profile can be obtained directly from disk amplitudes \cite{Billo:2002hm}.
In order to be self-contained, we briefly review this derivation.

We consider type IIB superstrings in the Euclidean space ${\mathbb R}^{10}$ (whose
coordinates we label by the indices $M,N=1,\ldots,10$) and place $N$ D3-branes
along the first four directions (labeled by the indices $\mu,\nu=1,\ldots,4$).
The six transverse directions are labeled instead by the indices $m,n=5,\ldots,10$. Under this
$\mathrm{SO}(10)\to \mathrm{SO}(4)\times \mathrm{SO}(6)$ decomposition,
the string coordinates $X^M$ and $\psi^M$ split as
\begin{equation}
X^M~\to~(X^\mu\,,\,X^m)~~~~{\rm
and}~~~~\psi^M~\to~(\psi^\mu\,,\,\psi^m)~~,
\label{xpsidec}
\end{equation}
while the anti-chiral spin fields $S^{\dot{\mathcal{A}}}$ ($\dot{\mathcal{A}}=1,\ldots,16$)
of the RNS formalism become products of four- and six-dimensional spin
fields according to
\begin{equation}
\label{spindec}
S^{\dot{\mathcal{A}}} ~\to~ (S_{\alpha}S_A\,,\,S^{\dot\alpha}S^{A})
\end{equation}
where the index $\alpha$ (or $\dot\alpha$) denotes positive (or negative)
chirality in four dimensions, and the upper (or lower) index $A$ indicates the
fundamental (or anti-fundamental) representation of
$\mathrm{SU}(4)\sim\mathrm{SO}(6)$.

The massless sector of the strings with both ends on D3-branes (3/3
strings) comprises the gauge field $A_\mu$, six scalars $\varphi_m$ and the
gauginos $\Lambda^{\alpha A}$ and $\bar\Lambda_{\dot\alpha A}$, which altogether form
the ${\cal N}=4$ vector multiplet. Their vertex operators are
\begin{equation}
\label{vertn4NS}
V_{A}(p)
= \frac{A_\mu(p)}{\sqrt 2} \,
\psi^{\mu}\,\ee^{-\phi} \,\ee^{\ii \,p \cdot X}~~~~,~~~~
V_{\varphi}(p)
= \frac{\varphi_m(p) }{\sqrt 2}\, \psi^{m}\,\ee^{-\phi}
\,\ee^{\ii\,p\cdot X}
\end{equation}
in the ($-1$) superghost picture of the NS sector, and
\begin{equation}
\label{vertn4R}
V_{\Lambda}(p)
= \Lambda^{\alpha A}(p)\,S_{\alpha}S_A\,\ee^{-\frac{1}{2}\phi}\,
\ee^{\ii\,
p\cdot X}~~~~,~~~~
V_{\overline\Lambda}(p)
= \overline\Lambda_{\dot\alpha A}(p)\,
S^{\dot\alpha}S^A\,\ee^{-\frac{1}{2}\phi}\, \ee^{\ii\, p\cdot X}
\end{equation}
in the ($-1/2$) picture of the R sector. Here $\phi$ is the boson of the superghost fermionization
formulas, $p$ is the longitudinal incoming momentum and
the convention $2\pi\alpha'=1$ has been taken.
The vertices (\ref{vertn4NS}) and (\ref{vertn4R}) describe fields in the adjoint representation of
$\mathrm{U}(N)$, and their scattering amplitudes give rise to the usual ${\cal N}=4$
SYM theory in the field theory limit $\alpha'\to 0$.

As is well-known, the D3-branes break half of the bulk supersymmetries in target space
due to the identification between left- and right-moving
spin fields enforced at the boundary, {\it i.e.}
\begin{equation}
\label{dottedbc}
S_\alpha(z) \,S_A(z) =-
\left.\widetilde S_\alpha(\overline z) \,\widetilde S_A(\overline z)\right|_{z=\overline z}
~~~~,~~~~
S^{\dot\alpha}(z) \,S^A(z) =
\left.\widetilde S^{\dot \alpha}(\overline z) \,\widetilde S^A(\overline z)
\right|_{z=\overline z}~.
\end{equation}
Let us now add the D($-1$) branes. They correspond to imposing Dirichlet
boundary conditions on all string coordinates $X^M$ and $\psi^M$, and
enforcing the following identification on the spin fields \cite{Billo:2002hm}
\begin{equation}
\label{undottedbc}
S_\alpha(z) \,S_A(z)
=\left.\widetilde S_\alpha(\overline z) \,\widetilde S_A(\overline z)\right|_{z=\overline z}
~~~~,~~~~
S^{\dot\alpha}(z) \,S^A(z)
=\left.\widetilde S^{\dot \alpha}(\overline z) \,\widetilde S^A(\overline z)
\right|_{z=\overline z}~~.
\end{equation}
By comparison with (\ref{dottedbc}), it is clear that the conditions (\ref{undottedbc}) break a
further half of the bulk supersymmetries, so that only eight
supercharges (those with spinor indices of the type $(\dot\alpha A)$) are preserved on
both branes.

The open strings with both ends on the D-instantons ($(-1)/(-1)$
strings) do not carry any momentum since there are no longitudinal Neumann directions.
Thus, these strings describe {\it moduli} rather than
dynamical fields. In the NS sector there are ten bosonic moduli corresponding to the
physical vertices
\begin{equation}
\label{vertA}
V_a = g_0\,{a'_\mu} \,\psi^{\mu}\,\ee^{-\phi}
~~~~,~~~~
V_\chi =\frac{\chi_m}{\sqrt 2} \,\psi^{m}\,\ee^{-\phi}~~,
\end{equation}
while in the R sector there are sixteen fermionic moduli whose
vertices are
\begin{equation}
\label{vertM'}
V_{M'}= \frac{g_0}{\sqrt{2}}\,{M'}^{\alpha A}\,
S_{\alpha}S_A\,\ee^{-\frac{1}{2}\phi}~~~~,~~~~
V_{\lambda}=
{{\lambda_{\dot\alpha A}}}\,S^{\dot\alpha}S^A
\,\ee^{-\frac{1}{2}\phi}~~.
\end{equation}
In writing the polarizations of these vertices we have adopted the traditional
notation; in particular we have distinguished the bosonic moduli into
four $a'_\mu$ (corresponding to the longitudinal directions of the D3 branes)
and six $\chi_m$ (corresponding to the transverse directions to the D3's).
Furthermore, $g_0$ is the dimensionful coupling for the effective theory on
the D-instantons, which is related to the Yang-Mills coupling on the D3 branes
by
\begin{equation}
\label{g0gym}
g_0 = \frac{g_{\mathrm{YM}}}{4\pi^2\alpha'}~~.
\end{equation}
Clearly, if $g_{\rm YM}$ is kept fixed when $\alpha'\to 0$ (as is appropriate to
retrieve the gauge theory on the D3-branes), then $g_0$ blows up.
Thus, as discussed in \cite{Billo:2002hm}, suitable factors of $g_0$, like the ones appearing
in (\ref{vertA}) and (\ref{vertM'}), must be present in the vertex operators
to retain non-trivial interactions when $\alpha'\to 0$.
As a consequence the moduli acquire non-trivial scaling dimensions which
turn out to be the right ones for their interpretation as parameters of an instanton solution
\cite{Dorey:2002ik,Billo:2002hm}. For instance, the
$a'_\mu$'s in (\ref{vertA}) have dimensions of (length) and are related to the positions of the
(multi)-centers of the instanton. Finally, we recall that since there are $k$
D-instantons, all the above moduli carry Chan-Paton factors of the adjoint
representation of $\mathrm{U}(k)$.

Let us now consider the open strings that are stretched
between a D3 and a D($-1)$ brane, {\it i.e.} the 3/($-1$) or ($-1$)/3 strings.
They are characterized by the fact that the four longitudinal directions along the D3 branes
have mixed Neumann-Dirichlet boundary conditions, while the
remaining six transverse directions have Dirichlet-Dirichlet
boundary conditions. As for the ($-1$)/($-1$) strings, also in
this case there is no momentum, and
the string excitations describe again moduli rather than
dynamical fields. In the NS sector the physical vertex operators
are
\begin{equation}
\label{vertexw}
V_w =
\frac{g_0}{\sqrt{2}}\,{w}_{\dot\alpha}\,
\Delta\, S^{\dot\alpha}\,\ee^{-\phi}~~~~,~~~~
V_{\overline w} =
\frac{g_0}{\sqrt{2}}
\,{\overline w}_{\dot\alpha}\, \overline\Delta\, S^{\dot\alpha}\, \ee^{-\phi}~~,
\end{equation}
where $\Delta$ and $\overline\Delta$ are the bosonic twist and anti-twist operators with
conformal weight $1/4$ which change the boundary conditions of the $X^\mu$ coordinates from
Neumann to Dirichlet and vice-versa by introducing a cut in the world-sheet \cite{Dixon:jw}.
The moduli ${w}_{\dot\alpha}$ and ${\overline w}_{\dot\alpha}$, whose $\mathrm{SO}(4)$ chirality
is fixed by the GSO projection, carry
Chan-Paton factors, respectively, in the bi-fundamental representations
$\mathbf{N}\times \mathbf{k}$ and $\mathbf{\overline N}\times \mathbf{\overline k}$ of the
gauge groups. Thus, one should write more explicitly $w_{\dot\alpha}^{\,~iu}$
and ${\overline w}_{\dot\alpha ui}$, where $u=1,\ldots,N$ and $i=1,\ldots,k$.
In the R sector of the mixed strings, the physical vertices are
\begin{equation}
\label{vertexmu}
V_\mu= \frac{g_0}{\sqrt{2}}\, {\mu}^{A}\,
\Delta\,S_{A}\, \ee^{-\frac 12 \phi}
~~~~,~~~~
V_{\overline\mu}= \frac{g_0}{\sqrt{2}}\, {{\overline \mu}}^{A}\,
\overline\Delta\,S_{A}\, \ee^{-\frac 12 \phi}
\end{equation}
where $\mu$ and $\overline \mu$ carry the same Chan-Paton factors as $w$ and $\overline w$
respectively. Again, it is the GSO projection, together with the
conserved supercurrent, that fixes the ${\rm SO}(6)$ chirality of the spin
fields in (\ref{vertexmu}).

The vertices described up to now exhaust the BRST
invariant spectrum of the open strings with at least one end point
on the D-instantons. However, to compute the couplings among the moduli
and derive the ADHM measure on moduli space from string interactions, it
is convenient to introduce also some \emph{auxiliary moduli} that
disentangle quartic interactions \cite{Billo:2002hm}.
In this context a particularly
relevant role is played by the auxiliary vertex
\begin{equation}
\label{vertDmod}
V_D= \frac 12 \, D_{\mu\nu}^{-} \psi^\nu\psi^\mu
~~,
\end{equation}
which describes an excitation of the ($-1$)/($-1$) strings
associated to an anti-self-dual tensor $D_{\mu\nu}^-=D_c^-\,\overline\eta_{\mu\nu}^c$
(where $\overline\eta_{\mu\nu}^c$ are the three anti-self-dual 't
Hooft symbols).

The transformation properties under the various groups and the
scaling dimensions of all ADHM moduli are summarized in
Table \ref{tab:1}.
\TABLE{
\label{tab:1}
\begin{tabular}{|c|c|c|c|c|c|}
\hline
 &$\mathrm{SO}(4)\simeq\mathrm{SU}(2)_+\times\mathrm{SU}(2)_-$&
 $\mathrm{SO}(6)\simeq \mathrm{SU}(4)$ & $\mathrm{U}(N)$ & $\mathrm{U}(k)$ & dimensions \\
\hline
$a'_\mu
$&$(\mathbf{2},\mathbf{2})$ & $\mathbf{1}$ & $\mathbf{1}$ &
 \bf{adj} & (length)$^1$ \\
 $\chi_m$&$(\mathbf{1},\mathbf{1})$ & $\mathbf{6}$ & $\mathbf{1}$ &
 \bf{adj} & (length)$^{-1}$ \\
 ${M'}^{\alpha A}$&$(\mathbf{2},\mathbf{1})$ & $\mathbf{4}$ & $\mathbf{1}$ &
 \bf{adj} & (length)$^{1/2}$ \\
 ${\lambda}_{\dot\alpha A}$&$(\mathbf{1},\mathbf{2})$ & $\mathbf{\overline 4}$ & $\mathbf{1}$ &
 \bf{adj} & (length)$^{-3/2}$ \\
 $ w_{\dot\alpha}$&$(\mathbf{1},\mathbf{2})$ & $\mathbf{1}$ & $\mathbf{N}$ &
 $\mathbf{k}$ & (length)$^{1}$ \\
 $ {\overline w}_{\dot\alpha}$&$(\mathbf{1},\mathbf{2})$ & $\mathbf{1}$ & $\mathbf{\overline N}$
  &$\mathbf{\overline k}$ & (length)$^{1}$ \\
 ${\mu}^{A}$&$(\mathbf{1},\mathbf{1})$ & $\mathbf{4}$ & $\mathbf{N}$
  &$\mathbf{k}$ & (length)$^{1/2}$ \\
  $ {\overline \mu}^A$&$(\mathbf{1},\mathbf{1})$ & $\mathbf{4}$ & $\mathbf{\overline N}$
  &$\mathbf{\overline k}$ & (length)$^{1/2}$ \\
$D_c^-$&$(\mathbf{1},\mathbf{3})$ & $\mathbf{1}$ & $\mathbf{1}$ & \bf{adj} &(length)$^{-2}$ \\
\hline
\end{tabular}
\caption{Transformation properties and scaling dimensions of the ADHM moduli.}
}

If we now compute all tree-level diagrams with insertions of the vertices listed above and
take the field theory limit $\alpha'\to 0$ (with $g_{\rm YM}$ fixed and hence
$g_0\to\infty$), we obtain the complete ADHM measure for
the instanton moduli space of the $\mathcal{N}=4$ SYM theory (see for instance Eq. (3.29)
in \cite{Billo:2002hm}).
An essential point is that the moduli $D_c^-$ and
$\lambda_{\dot\alpha A}$ appear in this measure as Lagrange multipliers, respectively, for the
bosonic and fermionic ADHM constraints. In particular, the bosonic constraints are the
following three $k\times k$ matrix equations
\begin{equation}
\label{bosADHM}
{W}^c +\ii\,
\overline\eta_{\mu\nu}^c\big[{a'}^\mu,{a'}^\nu\big] = \mathbf{0}~~,
\end{equation}
where
$(W^c)_j^{~i} = w_{\dot\alpha}^{\,~iu}\,(\tau^c)^{\dot\alpha}_{~\dot\beta}
\, \overline w^{\,\dot\beta}_{~uj}
$
in terms of the Pauli matrices $\tau^c$, while
the fermionic constraints are
\begin{equation}
\label{fermADHM}
{w}_{\dot\alpha}^{~u}\,\overline{\mu}_{u}^{\,A} +
{\mu}^{uA}\overline{w}_{\dot\alpha u} +
\big[a'_{\alpha\dot\alpha},{M'}^{\alpha A}\big] = \mathbf{0}~~.
\end{equation}

As explained in \cite{Billo:2002hm}, in the D3/D($-1$) system it is possible to consider
also disk diagrams with both mixed boundary conditions and insertions of
massless vertices of the 3/3 strings associated to gauge fields, and show
that on such mixed disks the various components of the gauge multiplet
may have non-trivial tadpoles and a non-vanishing space-time profile.
For example, for the vector field $A_\mu$ one finds indeed that
\begin{equation}
\big\langle\,{\cal V}_{A_\mu}\,\big\rangle_{\rm mixed~disk}
\not= 0
\label{vev4}
\end{equation}
where ${\cal V}_{A_\mu}$ is the gluon vertex $V_A$ defined in
(\ref{vertn4NS}) without polarization.
Furthermore, by taking the Fourier transform of these massless
tadpoles, after including a propagator and imposing the ADHM
constraints, one obtains \cite{Billo:2002hm} a space-time profile which is
precisely that of the classical gauge instanton solution in the singular gauge.
In other words, the D-instantons act as sources emitting non-abelian gauge fields.

\vskip 0.8cm
\section{D3/D(--1) systems in presence of a $B$-field}
\label{sec:d3d-1wb}

In this section we consider systems of D3 and D($-1)$ branes in presence
of a constant anti-symmetric tensor $B$ of the closed string NS-NS
sector, and focus in particular on the massless spectrum of the various kinds of open
strings to study their relation with the ADHM instanton construction
in non-commutative gauge theories.

The action for superstrings moving in a background $B$-field is
\footnote{For the fermionic part we use the action given in
\cite{Haggi-Mani:2000uc} which enjoys the property that the boundary terms in its variation can
be canceled by consistently imposing  on $\psi^M$ and $\delta\psi^M$ the {\it same}
constraints. See for instance \cite{Mihailescu:2000dn} for a discussion of the brane supersymmetry in
presence of $B$ field within the GS formalism.}
\begin{equation}
\label{action}
\begin{aligned}
S =&
-\frac{1}{4\pi\alpha'}\!\int\!d\sigma\,d\tau\Big\{\delta_{MN}\,\partial_a X^M \partial^a X^N +
\epsilon^{ab}B_{MN}\,\partial_a X^M\partial_b X^N
\Big\}+
\\
&
- \frac{\ii}{4\pi}\!\int\!d\sigma\,d\tau
\Big\{E_{MN}\,\overline\psi^M \partial\!\!\!/\,
\psi^N \Big\}~~.
\end{aligned}
\end{equation}
where $\epsilon^{\tau\sigma}=-\epsilon^{\sigma\tau}=1$, and
$E_{MN}=\delta_{MN}+B_{MN}$.
Varying $S$, we get a bosonic boundary term
\begin{equation*}
\int d\tau\Big[\delta X_M \left(\partial_\sigma X^M-B^M_{\phantom MN}
\partial_\tau X^N\right)\Big]^{\sigma=\pi}_{\sigma=0}~~,
\label{bcbos}
\end{equation*}
and a fermionic one
\begin{equation*}
\int d\tau\Big[E_{MN}\left(\psi^M_+\delta\psi^N_+-\psi^M_-
\delta\psi^N_-\right)\Big]^{\sigma=\pi}_{\sigma=0}
\label{bcfer}
\end{equation*}
where $\psi^M_\pm$ are the left and right-moving components of the world-sheet spinors $\psi^M$.
The boundary terms vanish after imposing boundary conditions of Dirichlet ({\bf D}) or
Neumann ({\bf N}) type on the open string fields.
For the bosonic coordinates we have
\begin{equation}
\mathbf{D}:~~\delta X^M\Big|_{\sigma=\bar\sigma}
=0\qquad\Rightarrow
\qquad\partial_\tau X^M\Big|_{\sigma=\bar\sigma}
\label{Dir boson}
=0
\end{equation}
or
\begin{equation}
\mathbf{N}:~~
\big(\partial_\sigma X^M-B^M_{\phantom MN}\partial_\tau X^N
\big)\Big|_{\sigma=\bar\sigma}
=0
\label{Neu boson}
\end{equation}
where $\bar\sigma=0$ or $\bar\sigma=\pi$.
For the fermionic fields, instead, the presence of $B$
requires some extra care. The Dirichlet boundary conditions are as usual
\begin{equation}
\label{Dir fermion}
\mathbf{D}:~~
\big(\delta\psi^M_++\eta_{\bar\sigma}\delta\psi^M_-\big)\Big|_{\sigma=\bar\sigma}=
\big(\psi^M_++\eta_{\bar\sigma}\psi^M_-\big)\Big|_{\sigma=\bar\sigma}=0
\end{equation}
where $\eta_{\bar\sigma}=\pm 1$, but there are two inequivalent ways of imposing
the Neumann boundary conditions, namely
\begin{subequations}
\label{Neu fermion}
\begin{align}
\label{Neu fermionA}
&\mathbf{N(a)}
:~
\big(E_{NM}\delta\psi^N_+-\eta_{\bar\sigma}
E_{MN}\delta\psi^N_-\big)
\Big|_{\sigma=\bar\sigma} =
\big(E_{NM}\psi^N_+-\eta_{\bar\sigma}  E_{MN}\psi^N_-\big)
\Big|_{\sigma=\bar\sigma}=0\\
\label{Neu fermionB}
&\mathbf{N(b)}
:~
\big(E_{MN}\delta\psi^N_+-\eta_{\bar\sigma}
E_{NM}\delta\psi^N_-\big)
\Big|_{\sigma=\bar\sigma}=
\big(E_{MN}\psi^N_+-\eta_{\bar\sigma} E_{NM}\psi^N_-\big)
\Big|_{\sigma=\bar\sigma}=0~~.
\end{align}
\end{subequations}
Clearly, if $B=0$ there is no distinction between
(\ref{Neu fermionA}) and (\ref{Neu fermionB}), but if $B\not =0$
they are different and thus
there will be various fermionic sectors when at
least one endpoint of the open string has boundary conditions of Neumann type.

In the following we will consider in detail a D3/D($-1$) system
with a constant background field $B$ along the four world-volume directions of the D3
branes, analyze the different kinds of open strings that are present
and make contact with non-commutative field theories and
the corresponding ADHM instanton construction.

\vskip 0.6cm
\subsection{The 3/3 strings}
\label{subsec:33}

In this sector the longitudinal coordinates $X^\mu$ and $\psi^\mu$
satisfy,
respectively, the boundary conditions of Neumann type (\ref{Neu boson}) and (\ref{Neu
fermion}) at both endpoints,
while the transverse coordinates $X^m$, $\psi^m$
satisfy, respectively, the Dirichlet boundary conditions (\ref{Dir boson}) and (\ref{Dir fermion})
at both endpoints.

After performing a Wick rotation on the world-sheet
($\tau \to -{\rm i}\tau_e$) and introducing the complex variable
$z=e^{\tau_e+\ii\sigma}$,
the bosonic boundary conditions may be written as
\begin{subequations}
\label{nbcz}
\begin{align}
\partial X^\mu(z,\overline z)&=
\Big(\frac{\uno+B}{\uno-B}\Big)^\mu_{\phantom\mu\nu}
\,\bar\partial  X^\nu(z,\overline z)~~,
\\
\partial X^m(z,\overline z)&=-\,\overline\partial  X^m(z,\overline
z)
\end{align}
\end{subequations}
for any $z\in \mathbb{R}$.
Following \cite{Chu:1999gi,Chu:1998qz}, we can solve the boundary
conditions (\ref{nbcz}) with the doubling trick by introducing  holomorphic chiral
bosons defined on the {entire} complex $z$-plane
\begin{equation}
X^M(z)=q^M-2\ii\alpha'p^M\log z+
\ii\sqrt{2\alpha'}\sum_{n\in\mathbb Z -\{0\}}\frac{\alpha^M_n}{n}\,z^{-n}~~,
\label{zmu}
\end{equation}
and writing
\begin{subequations}
\begin{align}
X^\mu(z,\overline z)&=
\frac12\Big[X^\mu(z)+\Big(\frac{\uno-B}{\uno+B}\Big)^{\!\mu}_{\phantom\mu\nu}\,
X^\nu(\overline z)\Big]
~~,
\label{33Xexp}
\\
X^m(z,\overline z)&=
x_0^m+\frac12\Big[X^m(z)-X^m(\overline z)\Big]
\label{33trans}
\end{align}
\end{subequations}
for any $z$ with ${\rm Im}(z)\geq 0$. In (\ref{33trans}) $x_0^m$ denotes the position of the D3
brane in the transverse space, which  can be set to zero without loss of generality.
Upon canonical quantization the
oscillators in (\ref{zmu}) become operators that satisfy the
following commutation relations
\begin{equation}
\label{zboscomm}
\big[q^\mu,q^\nu\big]=2\pi{\rm i}\,\alpha'B^{\mu\nu}
~~,~~\big[q^M,p^N\big]={\rm i}\,\delta^{MN}
~~,~~\big[\alpha^M_n,\alpha^N_n\big]=n\,\delta_{n+m,0}\,\delta^{MN}~~.
\end{equation}
The crucial difference with respect to the case at zero background is the non
trivial commutator among the longitudinal $q$'s that implies that the geometry on the
world-volume of the D3 brane
is non-commutative with a non-commutativity parameter~\footnote{It is worth
pointing out that the expression of the open string coordinates written in (\ref{33Xexp}) is
different from the one usually considered in the literature. In
particular, with our choice the open string metric is equal to
the closed string one ({\it i.e.} $\delta^{\mu\nu}$ in our case)
and the non-commutativity parameter $\theta$ is simply proportional to the background
field $B$ as shown in (\ref{theta}).
This is to be contrasted with the Seiberg-Witten approach \cite{Seiberg:1999vs} where a
different scaling is considered.
A discussion on the relation between these two approaches can be found for example in
Ref. \cite{Chu:2000pc}.}
\begin{equation}
\theta^{\mu\nu}=2\pi\alpha'B^{\mu\nu}
\label{theta}
\end{equation}
which is kept fixed in the field theory limit $\alpha'\to 0$.

Let us now consider the fermionic coordinates. As already
pointed out, when $B\not = 0$ there are two ways of imposing the boundary
conditions of Neumann type on the $\psi$'s and thus, in principle, there are four
different fermionic sectors. One possibility is to impose the
conditions (\ref{Neu fermionA}) at both endpoints for the
longitudinal directions, {\it i.e.}
\begin{subequations}
\label{bcfz}
\begin{align}
&\psi^{\mu}_{+}(z)= \eta_0 \Big(\frac{\uno+B}{\uno-B}\Big)^{\mu}_{\phantom\mu\nu}
\psi^{\nu}_{-}(\overline z)~~~~~~{\text{for}\,\,\, z\in\mathbb{R}_+}~~,
\\
&\psi^{\mu}_{+}(z)=\eta_{\pi} \Big(\frac{\uno+B}{\uno-B}\Big)^{\mu}_{\phantom\mu\nu}
\psi^{\nu}_{-}(\overline z)
~~~~~~{\text{for}\,\,\,z\in\mathbb{R}_-}~~,
\end{align}
\end{subequations}
and the conditions (\ref{Dir fermion}) for the transverse
directions, {\it i.e.}
\begin{subequations}
\label{bcfd}
\begin{align}
&\psi^m_{+}(z)= -\eta_0 \,\psi^m_{-}(\overline z)~~~~~~{\text{for}\,\,\, z\in\mathbb{R}_+}~~,
\\
&\psi^m_{+}(z)=-\eta_{\pi}\,\psi^m_{-}(\overline z)
~~~~~~{\text{for}\,\,\,z\in\mathbb{R}_-}~~.
\end{align}
\end{subequations}
As usual, if $\eta_0\eta_\pi=1$ we obtain the R sector, while if
$\eta_0\eta_\pi=-1$ we obtain the NS sector.
The boundary constraints (\ref{bcfz}) and (\ref{bcfd}) can be
solved by introducing holomorphic fermionic fields defined on the
entire complex plane such
that
\begin{equation}
\psi^M({\rm e}^{2\pi\ii}\,z)= -\eta_0\eta_\pi\, \psi^M(z)~~,
\label{monodromy}
\end{equation}
and then by writing
\begin{subequations}
\begin{align}
&\psi^\mu_+(z)=z^{\frac12}\psi^\mu(z)~~,~~
\psi^\mu_-(\overline z)=\eta_0\,{\overline z}^{\frac12}
{\Big(\frac{\uno-B}{\uno+B}\Big)}^{\!\mu}_{\phantom\mu\nu}\,\psi^\nu (\overline z)
~~,\label{33fermlong}\\
&\psi^m_+(z)=z^{\frac12}\psi^m(z)~~,~~
\psi^m_-(\overline z)=-\eta_0\,{\overline z}^{\frac12}
\,\psi^m (\overline z)
\label{33fermtran}
\end{align}
\end{subequations}
for any $z$ with ${\rm Im}(z)\geq 0$. {}From (\ref{monodromy}) it
easily follows that
\begin{equation}
\psi^M(z) = \sum_{r\in \mathbb{Z}+\nu} \psi^M_r\,z^{-r-1/2}
\label{psiz}
\end{equation}
where $\nu=0$ in the R sector and $\nu=1/2$ in the NS sector.
Upon canonical quantization the fermionic
modes in (\ref{psiz}) become operators that satisfy the
standard anti-commutation relations
\begin{equation}
\big\{\psi^M_r,\psi^N_s\big\}=\delta_{r+s,0}\,\delta^{MN}~~.
\label{psipsi}
\end{equation}

{}From (\ref{zboscomm}) and (\ref{psipsi}) we clearly see that the
excitation spectrum of these open strings is isomorphic to the one
of the 3/3 strings without the $B$ background.
In particular at the massless level we find a gauge field $A_\mu$
and six scalars $\phi_m$, together with their fermionic partners
$\Lambda^{\alpha A}$ and ${\overline \Lambda}_{\dot \alpha A}$
that complete a $\mathcal{N}=4$ vector supermultiplet in the adjoint
representation of $\mathrm{U}(N)$. The corresponding vertex
operators have the same expressions as those listed in section
\ref{sec:d3d-1}, see in particular Eqs. (\ref{vertn4NS}) and (\ref{vertn4R}).
However, since the longitudinal zero-modes $q^\mu$'s contained in the
exponential ${\rm e}^{\ii p\cdot X}$
satisfy non-trivial commutation relations, the scattering
amplitudes among these vertices are modified and the resulting gauge theory
becomes non-commutative \cite{Seiberg:1999vs}.
Typically, in the field theory limit where the non-commutative parameter $\theta$ defined in
(\ref{theta}) is kept fixed, various interaction terms may acquire momentum factors like
$\cos (p_1\wedge p_2)$ and $\sin(p_1\wedge p_2)$ where
\begin{equation}
\label{wedge}
p_1\wedge p_2 = \frac 12 \,p_1^\mu \,\theta_{\mu\nu}\, p_2^\nu~~.
\end{equation}
Furthermore, new structures
may appear as well (see for instance Ref. \cite{Armoni:2000xr,Bonora:2000ga}). For instance in the 3-gluon vertex the usual term proportional to the
structure constants of $\mathrm{U}(N)$ is modified with $\cos(p_i\wedge p_j)$
factors and a term proportional
to the $d_{\hat a \hat b \hat c}$ tensor shows up in the non-commutative case
(see Fig. \ref{fig:ncv}).

\FIGURE{\centerline{
\psfrag{Amu}{\scriptsize ${\hat a},\mu$}
\psfrag{Bnu}{\scriptsize ${\hat b},\nu$}
\psfrag{Crho}{\scriptsize ${\hat c},\rho$}
\psfrag{k}{\scriptsize $p_1$}
\psfrag{p}{\scriptsize $p_2$}
\psfrag{q}{\scriptsize $p_3$}
\psfrag{vertex}{\small $V^{\mu\nu\rho}_{\hat a\hat b\hat c}(p_1,p_2,p_3)=
\left[(p_3-p_2)^\mu \delta^{\nu\rho}
+ (p_2-p_1)^\rho \delta^{\mu\nu} + (p_1-p_3)^\nu \delta^{\rho\mu}\right]$}
\psfrag{vertex1}{\small $\phantom{V^{\mu\nu\rho}_{\hat a\hat b\hat
c}(p_1,p_2,p_3)}\times\left[f_{\hat a\hat b\hat c}\cos(p_1\wedge p_2) + d_{\hat a\hat
b\hat c} \sin (p_1\wedge p_2)\right]$}
\hskip -10cm
\includegraphics[width=4.5cm]{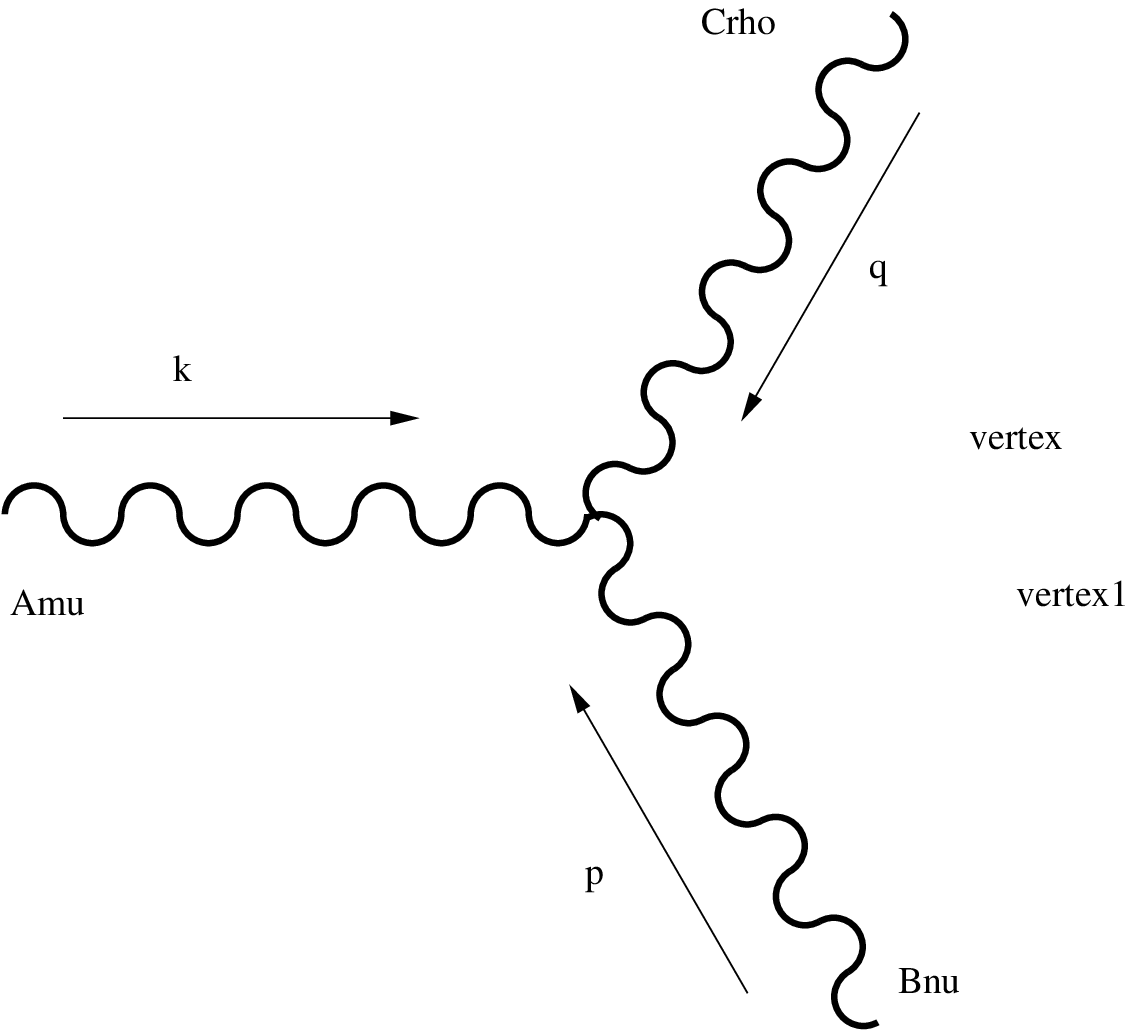}
\caption{\small The three-gluon vertex in the non-commutative $\mathrm{U}(N)$ Yang-Mills theory.}
\label{fig:ncv}
}}

However, there are other ways of imposing the Neumann boundary conditions on the fermionic
fields. For example, we could require that both endpoints of the open string satisfy
conditions of type (\ref{Neu fermionB}). In this case, essentially
nothing changes with respect to what discussed above. In fact, to solve these boundary conditions
one still introduces chiral fermions ${\psi'}^M(z)$ with the same monodromy
properties, and hence the same mode expansion, of the fields $\psi^M(z)$
defined in (\ref{psiz}). Therefore, the resulting spectrum is simply a
copy of the one previously considered, and in particular at the
massless level we find a gauge vector multiplet.
If instead we impose the boundary conditions (\ref{Neu fermionA}) at $\sigma=0$ and
the conditions (\ref{Neu fermionB}) at $\sigma=\pi$, or vice-versa,
things are radically different.
In fact, to solve the corresponding constraints we have to
introduce chiral fermions $\chi^\mu(z)$ such that
\begin{equation*}
\chi^\mu({\rm e}^{2\pi\ii}\,z)=-\eta_0\eta_\pi
\Big[\Big(\frac{\uno\pm B}{\uno\mp
B}\Big)^{\!2}\,\Big]^\mu_{\phantom\mu\nu}\,
\chi^\nu(z)~~.
\end{equation*}
These fields are no longer periodic or anti-periodic, and hence their
modes are no longer integers or half-integers. Moreover, it can be
checked that the physical spectrum constructed using these modes
does not contain massless states even in the field theory limit.
In particular it is not possible to obtain a massless gauge vector
with this ``mixed'' choice of fermionic boundary
conditions. Therefore, for our purposes such strings do not play any role
and can be consistently neglected in the classical approximation.

In conclusions, the 3/3 strings have only two sectors that in the field theory limit
reproduce a non-commutative $\mathcal N=4$ SYM theory. To just describe
this gauge theory it would be sufficient to consider only one of them, as
is usually done in the literature. However, as we shall see later, to obtain also
the non-commutative ADHM instanton construction from string theory
it is necessary to consider both sectors in a symmetric way, {\it
i.e.} identify states with equal quantum numbers.
This implies, for instance,
that the gluon emission vertex in the ($-1$) superghost
picture is
\begin{equation}
\label{glu}
V_{A}(p)  = \frac{{A_\mu(p)}}{\sqrt{2}}\,
\frac{\psi^{\mu}+{\psi'}^\mu}{\sqrt{2}}\,\ee^{-\phi} \,
\ee^{\ii p \cdot X}~~.
\end{equation}
However, in all practical calculations we can simply
identify $\psi^\mu$ and ${\psi'}^\mu$, and still use the properly
normalized gluon vertex (\ref{vertn4NS}) and the standard contraction rules.

\vskip 0.6cm
\subsection{The (--1)/(--1) strings}
\label{subsec:11}

The open strings with both endpoints on the D$(-1)$ branes have Dirichlet boundary
conditions (\ref{Dir boson}) and (\ref{Dir fermion}) in all
directions and hence do not feel any effect of the $B$ background.
All bosonic coordinates $X^M$ have an expansion like (\ref{33trans})
with $x_0^M$ denoting the position of the D-instantons, while all
fermionic coordinates $\psi^M$ are as in (\ref{33fermtran}).
The physical spectrum of these (--1)/(--1) strings contains the
same states as in the free case, describing the moduli
$a'_\mu$, $\chi_m$, ${M'}^{\alpha A}$ and $\lambda_{\dot\alpha
A}$ together with the auxiliary fields $D_c^-$. Their corresponding
vertices have the same expressions as in (\ref{vertA}),
(\ref{vertM'}) and (\ref{vertDmod}).

\vskip 0.6cm
\subsection{The (--1)/3 and 3/(--1) strings}
\label{subsec:31}

We now consider the open strings that stretch between a D($-1$) and a D3
brane. To simplify our discussion, but without loosing generality,
we assume that the background field $B_{\mu\nu}$ is in the skew-diagonal
form
\begin{equation}
\label{Bskew}
B=\begin{pmatrix}
\begin{matrix}0 & b^{\dot 2}\\ -b^{\dot 2}& 0\end{matrix} &\mathbf{0} \\
\mathbf{0} & \begin{matrix}0 & b^{\dot 1}\\ -b^{\dot 1}& 0\end{matrix}
\end{pmatrix}~~,
\end{equation}
so that it becomes natural to introduce the complex fields
\begin{subequations}
\begin{align}
\label{zI}
& Z^{\dot 1} = \frac{X^3 + \ii\, X^4}{\sqrt{2}}~~,
\hskip 0.8cm
Z^{\dot 2} = \frac{X^1 + \ii\, X^2}{\sqrt{2}}~~,\\
&
\label{PsiI}
\Psi^{\dot 1} = \frac{\psi^3 + \ii\, \psi^4}{\sqrt{2}}~~,
\hskip 1cm
\Psi^{\dot 2} = \frac{\psi^1 + \ii\, \psi^2}{\sqrt{2}}~~.
\end{align}
\end{subequations}
As we will see in the following, the use of dotted indices,
like for anti-chiral spinors, turns out to be particularly useful.
In the above complex basis the bosonic boundary conditions for the longitudinal coordinates of a
($-1$)/3 string
become
\begin{subequations}
\label{bc3-1z}
\begin{align}
\label{bc3-1zd}
\partial Z^{\dot\alpha}(z,\overline z) & = -\overline\partial Z^{\dot\alpha}(z,\overline z)
~~~~~~ ~~~~~~ ~~~~~~{\text{for}\,\,\, z\in\mathbb{R}_+}~~,
\\
\label{bc3-1zn}
\partial Z^{\dot\alpha}(z,\overline z) & =\left(
\frac{1 - \ii b^{\dot\alpha}}{1 + \ii b^{\dot\alpha}}\right)
\overline\partial Z^{\dot\alpha}(z,\overline z) ~~~~~~{\text{for}\,\,\,
z\in\mathbb{R}_-}~~,
\end{align}
\end{subequations}
where $\dot\alpha={\dot 1},{\dot 2}$.
Since the boundary conditions (\ref{bc3-1z}) are diagonal in each
complex direction, {\it i.e.} for a given value of $\dot\alpha$, for simplicity
we temporarily suppress this index
and reinstate it only when necessary.

To solve (\ref{bc3-1z}) we use again
the doubling trick: we introduce a complex chiral field $Z(z)$ such that
\begin{equation}
Z({\rm e}^{2\pi\ii}\,z)= -\left(\frac{1-\ii b}{1+\ii b}\right) Z(z)
=-
{\rm e}^{-2\pi\ii\epsilon}\,Z(z)
\label{monodromyZ}
\end{equation}
where
\begin{equation}
\label{def_epsi}
\epsilon =\frac{1 }{\pi}\arctan b\hskip 1cm
\Big(-\frac12<\epsilon<\frac{1}{2}\,\,\Big)~~,
\end{equation}
and write
\begin{equation}
Z(z,\overline z) = \frac{1}{2}\Big[Z(z)-\overline Z(\overline
z)\Big]
\label{Zopen}
\end{equation}
for any $z$ with ${\rm Im}(z)\geq 0$. {}From (\ref{monodromyZ}) we
deduce that
\begin{equation}
Z(z)=\sum_{n\in \mathbb{Z}}
\frac{1}{n+\epsilon+\frac{1}{2}}\,\alpha_{n+\epsilon+\frac{1}{2}}\,
z^{-n-\epsilon-\frac{1}{2}}~~,
\label{modeZ}
\end{equation}
which, for vanishing background,
reduces to the standard half-integer mode expansion of a
boson with mixed Dirichlet-Neumann boundary conditions. Canonical
quantization leads to the commutators
\begin{equation}
\left[\alpha_{n+\epsilon+\frac 12},
\overline\alpha_{-m-\epsilon-\frac 12}\right]=
\Big(n+\epsilon+\frac 12\Big)\,\delta_{n, m}~
\end{equation}
where $\overline\alpha_{-m-\epsilon-\frac{1}{2}}$ are the
modes that appear in the expansion of the complex
conjugate field $\overline Z$. The oscillators
with positive index are annihilation operators with respect to the
twisted vacuum $|\epsilon\rangle$, namely
\begin{equation*}
\label{vac1}
\alpha_{n+\epsilon+\frac 12}\ket{\epsilon}= 0~~ ~~ (n\ge 0)~~ ~~ ~~ {\rm and}~~ ~~ ~~
\overline\alpha_{n - \epsilon -\frac 12}\ket{\epsilon} = 0~~~~
(n\ge 1)~~,
\end{equation*}
whereas the modes with negative index are creation operators.
The contribution of this twisted boson to the Virasoro generator $L_0$
is
\begin{equation*}
\label{L0Z}
L_0^{(Z)}=
\sum_{n\in \mathbb{Z}} \,:\!\overline\alpha_{-n -\epsilon - \frac 12}
\alpha_{n + \epsilon + \frac 12}\!:\,
+ \left(\frac 18  - \frac{\epsilon^2}{2}\right)
\end{equation*}
where the normal ordering is defined with respect to the twisted vacuum
introduced above. Thus, $|\epsilon\rangle$ has conformal dimension
$h= 1/8 - \epsilon^2/2$ and is created from the $\mathrm{SL}(2,\mathbb R)$
invariant vacuum $|0\rangle$ by a {twist field}
$\sigma(z)$ of weight $h$, namely
\begin{equation*}
\ket{\epsilon}=\lim_{z\rightarrow 0} \sigma(z)\ket0
~~.
\label{twist}
\end{equation*}

Let us now consider the longitudinal fermionic coordinates (\ref{PsiI}), for
which the boundary conditions are
\begin{subequations}
\label{bc3-1psi}
\begin{align}
\label{bc3-1psid}
\Psi^{\dot\alpha}_+(z) & = -\eta_0 \Psi^{\dot\alpha}_-(\overline z)
~~~~~~~~~~~~~~~~~~\,{\text{for}\,\,\, z\in\mathbb{R}_+}~~,
\\
\label{bc3-1psin}
\Psi^{\dot\alpha}_+(z) & = \eta_\pi \left(
\frac{1 - \ii b^{\dot\alpha}}{1 + \ii b^{\dot\alpha}}\right)
\Psi^{\dot\alpha}_-(\overline z)~~~~~~{\text{for}\,\,\, z\in\mathbb{R}_-}
\end{align}
\end{subequations}
if we choose the form (\ref{Neu fermionA}) of the Neumann
relation at $\sigma=\pi$. As before, we solve these constraints using the doubling
trick: for each value of the index $\dot\alpha$ we introduce a multi-valued chiral fermion
$\Psi(z)$ such that
\begin{equation}
\label{polPsi}
\Psi({\rm e}^{2\pi\ii}\,z)= \eta_0\eta_\pi\,
{\rm e}^{-2\pi\ii\epsilon}\,\Psi(z)
\end{equation}
and write
\begin{equation}
\Psi_+(z)=z^{\frac12}\Psi(z)~~~~,~~~~
\Psi_-(\overline z)=-\eta_0\,{\overline z}^{\frac12}
\,\Psi(\overline z)
\end{equation}
for any $z$ with ${\rm Im}(z)\geq 0$.
{}From the monodromy property (\ref{polPsi}) we easily find that
\begin{equation}
\label{modePsi}
\Psi(z) = \sum_{n\in \mathbb{Z}+\nu} \Psi_{n+\epsilon} z^{-n - \epsilon -\frac 12}
\end{equation}
where $\nu=0$ in the NS sector ($\eta_0\eta_\pi=-1$) and $\nu=1/2$
in the R sector ($\eta_0\eta_\pi=1$).
Canonical Dirac quantization leads to the following
non-vanishing anti-commutators
\begin{equation}
\label{fc}
\Big\{\Psi_{n+\epsilon},\overline \Psi_{-m-\epsilon}\Big\} =\delta_{n,m}
\end{equation}
where $\overline \Psi_{-m-\epsilon}$ are the modes of the complex conjugate field
$\overline\Psi$. Notice that in the presence of a $B$ field,
neither the NS nor the R sectors of the mixed directions have
zero-modes, and thus for the ($-1$)/3 strings the twisted fermionic vacuum
$|\epsilon\rangle$, annihilated by all positive modes, is always non degenerate.
The contribution of $\Psi$ to the Virasoro operator $L_0$ is
\begin{equation}
L_0^{(\Psi)}  =\sum_{n\in \mathbb{Z}+\nu} (n+\epsilon)
\,:\!\overline \Psi_{-n-\epsilon} \Psi_{n+\epsilon}\!:\,
+ a_\nu
\label{L0ferm}
\end{equation}
where the normal ordering constant is
\begin{equation}
\label{anu}
a_0 =\frac 12 \left(\frac 12 - |\epsilon|\right)^2~~~~,~~~~
a_{\frac 12}= \frac {\epsilon^2}{2}
\end{equation}
in the NS and R sectors respectively.

The twisted vacuum of the NS sector $|\epsilon\rangle_{\rm NS}$, whose energy
is $a_0$, is created from the $\mathrm{SL}(2,\mathbb{R})$
invariant vacuum $|0\rangle$ by the spin-twist field $s^+(z)$ when $\epsilon >0$
and by $s^-(z)$ when $\epsilon < 0$,
namely
\begin{equation*}
|\epsilon\rangle_{\rm NS}=
\left\{
\begin{array}{c}
\stackrel{\displaystyle{\lim}}{\scriptstyle{z\rightarrow 0}}
\,\,\stackrel{\displaystyle{s^+(z)\ket0~~~~{\rm
for}~~\epsilon>0~~,}}{}
\\
\stackrel{\displaystyle{\lim}}{\scriptstyle{z\rightarrow 0}}
\,\,\stackrel{\displaystyle{s^-(z)\ket0~~~~{\rm
for}~~\epsilon<0~~.}}{}
\end{array}
\right.
\end{equation*}
The spin-twist fields $s^\pm$ are most easily described in the bosonization formalism where
\begin{equation}
\label{bosPsi}
\Psi(z)= \ee^{+\ii \varphi(z)}~~~~,~~~~
\overline\Psi(z)= \ee^{-\ii \varphi(z)}
\end{equation}
up to cocycle terms \footnote{As explained in appendix \ref{app:a}, complex conjugation
acts as $\big(\Psi^{\dot\alpha}\big)^*=\overline \Psi_{\dot\alpha}$.}. Then, one can show that
\begin{equation}
\label{spin_NS}
s^\pm(z) = \ee^{\pm\ii \left(\frac 12 -
|\epsilon|\right)\varphi(z)}~~.
\end{equation}
In the following we will need to consider also the first excited
state of the NS sector. If $\epsilon>0$, this is
\begin{equation*}
\label{exc1}
\overline\Psi_{-\epsilon}\ket\epsilon_{\rm NS} =
\lim_{z\to 0}\overline t^+(z)\ket 0_{\rm NS}
\end{equation*}
where $\overline t^+(z)$ is an excited spin-twist field defined by the OPE
\begin{equation*}
\label{excsf2}
\overline\Psi(z) s^+(w) =
\frac{\overline t^+(w)}{(z-w)^{\frac 12 - \epsilon}} + \cdots ~~.
\end{equation*}
In the bosonization formalism, this OPE allows us to
write $\overline t^+(z)= \ee^{-\ii(\frac 12 + \epsilon)\varphi(z)}$
whose conformal dimension is $a_0+\epsilon=\frac 12(\frac 12 + \epsilon)^2$.
For $\epsilon < 0$, instead, the first excited state is
\begin{equation*}
\label{exc2}
\Psi_{\epsilon}\ket\epsilon_{\rm NS} =
\lim_{z\to 0}\, t^-(z)\ket 0
\end{equation*}
where $t^-(z) = \ee^{\ii(\frac 12 -\epsilon)\varphi(z)}$
whose conformal dimension is $a_0-\epsilon=\frac 12(\frac 12 - \epsilon)^2$.

In the R sector, the twisted vacuum $|\epsilon\rangle_{\rm R}$
is created from the $\mathrm{SL}(2,\mathbb{R})$
invariant vacuum by the spin-twist field
\begin{equation*}
\label{spin_R}
s_{\rm R}(z)=e^{-\ii\epsilon\varphi(z)}
\end{equation*}
whose conformal dimension is $\epsilon^2/2$.
For our future
applications we will not need to consider excited states in the
R sector, but of course they could be easily constructed along the
same lined discussed for the NS case.

It is important to realize that if we had chosen the other type of Neumann
boundary conditions for the longitudinal fermionic coordinates,
{\it i.e.} (\ref{Neu fermionB}), we would have retrieved the same
expressions as above at all stages, but with $b\to -b$, or equivalently with
$\epsilon\to -\epsilon$.
Thus, we can conclude that for
the ($-1$)/3 strings, the two possible choices of fermionic Neumann boundary conditions
are simply related to each other by the exchange of $\Psi$ and $\overline\Psi$.

\vskip 0.6cm
\subsubsection{Spectrum}
\label{subsubsec:spectrum31}

Let us now discuss the physical spectrum of the mixed
strings. Due to the absence of momentum in all directions, there
are very severe constraints on the form of allowed states and
only very few of them are physical.

In the NS sector the twisted vacuum $\ket{\epsilon_1;\epsilon_2}_{\rm NS}$
cannot be physical. Let us see why. If, for example,
$\epsilon_1,\epsilon_2>0$,
the vacuum is described by the following vertex operator in the ($-1$)-superghost picture
\begin{equation*}
\sigma_1\, s^+_1\,\sigma_2\, s^+_2 \,\ee^{-\phi}~~;
\label{vacuumvert}
\end{equation*}
if $\epsilon_1$ or $\epsilon_2$ are negative, the corresponding twist fields
$s^+$ must be replaced by $s^-$. The
conformal dimension of any of these vertices is $h=1-(|\epsilon_1|+|\epsilon_2|)/2$.
Thus, $h$ can never be 1 in a non-trivial background. Physical states can instead be present
in the first excited level. When the excitation is produced by
the longitudinal fermions, there are these possibilities
\begin{equation*}
\overline\Psi_{\dot 1\,,\,-\epsilon_1}\ket{\epsilon_1;\epsilon_2}_{\rm NS}
~~,~~
\Psi^{\dot 1}_{\epsilon_1}\ket{\epsilon_1,\epsilon_2}_{\rm NS}
~~,~~
\overline\Psi_{\dot 2\,,\,-\epsilon_2}\ket{\epsilon_1;\epsilon_2}_{\rm NS}
~~,~~
\Psi^{\dot 2}_{\epsilon_2}\ket{\epsilon_1,\epsilon_2}_{\rm NS}
\label{excitstates}
\end{equation*}
depending on whether $\epsilon_1 >0$, $\epsilon_1 <0$, $\epsilon_2 >0$ or $\epsilon_2
<0$. If, for example, $\epsilon_1,\epsilon_2 >0$, we have the following two vertex operators
(in the $(-1)$ superghost picture)
\begin{equation}
\label{vertexw_0}
\mathcal{V}_{\dot 1}=\sigma_1\,\overline t^+_1\,\sigma_2\, s^+_2 \,\ee^{-\phi}
~~~~{\rm and}~~~~
\mathcal{V}_{\dot 2}=\sigma_1 \,s^+_1\,\sigma_2\, \overline t^+_2 \,\ee^{-\phi}
~~.
\end{equation}
Again, if either $\epsilon_1$ or
$\epsilon_2$ are negative, we must replace $s^+$ and
$t^+$ with $s^-$ and $t^-$ in the appropriate places.
The total conformal weight of these vertices is
\begin{equation*}
\label{weight_w}
h\left(\mathcal{V}_{\dot 1}\right)  = 1 + \frac{|\epsilon_1| - |\epsilon_2|}{2}~~~~,~~~~
h\left(\mathcal{V}_{\dot 2}\right)  = 1 - \frac{|\epsilon_1| - |\epsilon_2|}{2}
\end{equation*}
and thus they are physical only if $|\epsilon_1|= |\epsilon_2|$,
{\it i.e.} when the background is self-dual or anti-self-dual.
Extending this analysis, one can easily prove
that no other physical states exist in the NS
sector.

In the R sector the only physical state turns out to be the vacuum,
which carries indices of the spinor representation of $\mathrm{SO}(6)$
due to the zero-modes of the $\psi^m$ fields in the transverse
directions. This vacuum is thus associated to the transverse spin fields
$S_A$ or $S^A$, and
the corresponding vertex operators in the $(-1/2)$ superghost picture are
\begin{equation}
\label{vmu}
\mathcal{V}_A = \sigma_1 \,s_{R,1}\, \sigma_2\, s_{R,2}\, S_A\,
\ee^{-\frac 12 \phi}~~~~{\rm and}~~~~
\mathcal{V}^A = \sigma_1 \,s_{R,1}\, \sigma_2 \,s_{R,2}\, S^A\,
\ee^{-\frac 12 \phi}
\end{equation}
which have conformal weight 1, because the $\epsilon$ contributions cancel between the bosonic
and the fermionic terms. Thus, the vertices (\ref{vmu}) describe physical states.
No other (excited) state of the R sector is physical.

\vskip 0.6cm
\subsubsection{Moduli spectrum in (anti-)self-dual background}
\label{subsubsec:asdbkg}

The previous discussion shows that the physical NS sector is non-empty
only when the background has a definite duality. For definiteness, let us
consider a self-dual $B$ field with $\epsilon_1=\epsilon_2 = \epsilon>0$.
In this case the physical NS vertices are given by (\ref{vertexw_0})
and in the bosonized formalism their fermionic parts read
\begin{equation}
\label{vwf}
\mathcal{V}_{\dot 1}
\sim
\ee^{-\frac{\ii}{2}(\varphi_1 - \varphi_2) -\ii\epsilon(\varphi_1 +
\varphi_2)}
~~~~{\rm and}~~~~
\mathcal{V}_{\dot 2}
\sim
\ee^{+\frac{\ii}{2}(\varphi_1 - \varphi_2)-
\ii\epsilon(\varphi_1 + \varphi_2)}
~~.
\end{equation}
We denote collectively these vertices
by $\mathcal{V}_{\dot\alpha}$ with $\dot\alpha=\dot 1, \dot 2$,
since they are created by the action of
$\overline\Psi_{\dot\alpha\,,\,-\epsilon}$ on the twisted
vacuum.
The label $\dot\alpha$ suggests that they transform as an anti-chiral
spinor of $\mathrm{SO}(4)$.
To prove this, let us write
the $\mathrm{SO}(4)$ generators as
\begin{equation}
\label{ferm_curr}
J_{\mu\nu}\equiv \,:\!\psi_\mu\psi_\nu\!: \,\,=\,\,\eta^c_{\mu\nu} J_c^{(+)} +
\overline\eta^c_{\mu\nu} J_c^{(-)}
\end{equation}
where $J_c^{(+)}$ and $J_c^{(-)}$ are the $\mathrm{SU}(2)_+$ and
$\mathrm{SU}(2)_-$ currents respectively, and then use the
bosonized formalism to get (up to cocycles)
\begin{subequations}
\label{ferm_curr_bos}
\begin{align}
\label{j+}
 &J^{(+)}_3 = \frac 12\,\big(\partial\varphi_1 + \partial\varphi_2\big)~~,~~
 J^{(+)}_\pm = \ii\,\ee^{\pm\ii(\varphi_1+\varphi_2)}~~,
\\
\label{j-}
 &J^{(-)}_3 = \frac 12\,\big(\partial\varphi_1 - \partial\varphi_2\big)~~,~~
 J^{(-)}_\pm = \ii\,\ee^{\mp\ii(\varphi_1-\varphi_2)}~~,
\end{align}
\end{subequations}
where $J^{(\pm)}_{\pm} = \big(J^{(\pm)}_1 \pm \ii J^{(\pm)}_2\big)$ are the step
operators of the $\mathrm{SU}(2)_\pm$ groups. In a
self-dual skew-diagonal background field $B_{\mu\nu}$,
the Lorentz group is broken to
$\mathrm{U}(1)_+\times\mathrm{SU}(2)_-$, where $\mathrm{U}(1)_+$ is the
subgroup of $\mathrm{SU}(2)_+$ generated by $J^{(+)}_3$.
Using (\ref{vwf}) and (\ref{ferm_curr_bos}), it is elementary to obtain
\begin{equation}
\label{opeJ-}
J^{(-)}_c(z) \,\mathcal{V}_{\dot\alpha}(w) =
\frac{\ii}{2}\,\frac{\mathcal{V}_{\dot\beta}(w)\,(\tau_c)^{\dot\beta}_{~\dot\alpha}}{z-w}
+ \cdots
\end{equation}
which shows that indeed the vertices $\mathcal{V}^{\dot\alpha}$
transform as a doublet of $\mathrm{SU}(2)_-$, and
\begin{equation}
\label{chargew}
J^{(+)}_3(z) \,\mathcal{V}_{\dot\alpha}(w) = \ii\,\frac{
\epsilon\,\mathcal{V}_{\dot\alpha}(w)}{z-w} + \cdots
\end{equation}
which shows that these vertices have charge $\epsilon$
under $\mathrm{U}(1)_+$~\footnote{Correctly, no simple
poles appear in the OPE of $\mathcal{V}^{\dot\alpha}$
with the broken generators $J^{(+)}_\pm$ of $\mathrm{SU}(2)_+$.}.
Due to this non-zero charge, the vertices $\mathcal{V}_{\dot\alpha}$
are {\it not} true anti-chiral spinors and, contrarily to naive expectations, they {\it cannot}
be associated with the moduli $w_{\dot\alpha}$ of the ADHM construction, which,
as shown in Table \ref{tab:1}, are singlets under $\mathrm{SU}(2)_+$ and
hence carry zero charge under $\mathrm{U}(1)_+$. Notice that when
$B=0$ the physical NS vertex operators (\ref{vertexw}) have the
quantum numbers of the degenerate twisted NS vacuum; on the contrary when
$B\not=0$ the physical NS vertices (\ref{vertexw_0})
are associated to the first fermionic
excited states on a non-degenerate scalar vacuum and hence carry the quantum numbers of the
fermionic oscillators, which are Lorentz vectors. This explains the
origin of the non-vanishing charge of $\mathcal{V}_{\dot\alpha}$
under $\mathrm{U}(1)_+$.

This problem can be overcome thanks to the existence of
another way of realizing the ($-1$)/3 strings. So far, in fact, we have
used fermionic fields $\Psi^{\dot\alpha}$ and $\overline\Psi_{\dot\alpha}$
that satisfy the Neumann boundary conditions
of type {\bf N(a)} (see Eq. (\ref{bc3-1psin})). However,
also the boundary conditions of type {\bf N(b)}
can be used. With this second choice, everything goes formally as before
except that in the fermionic sector $\epsilon$ is everywhere replaced by $- \epsilon$
and the roles of $\Psi^{\dot\alpha}$ and
$\overline\Psi_{\dot\alpha}$ are exchanged. Thus, in the new NS
sector, for a self-dual background with $\epsilon>0$, the physical
vertex operators are
\begin{equation}
\label{vertexw_0p}
{\mathcal{V}'}^{\dot 1}=\sigma_1\, t^-_1\,\sigma_2 \,s^-_2\, \ee^{-\phi}
~~~~{\rm and}~~~~{\mathcal{V}'}^{\dot 2}=\sigma_1\, s^-_1\,\sigma_2 \,t^-_2\, \ee^{-\phi}
\end{equation}
instead of the ones given in (\ref{vertexw_0}).
Computing their OPE's with the preserved Lorentz generators one
finds
\begin{subequations}
\label{ope1}
\begin{align}
\label{ope1-}
&J^{(-)}_c(z) \,{\mathcal{V}'}_{\dot\alpha}(w) =
\frac{\ii}{2} \,\frac{{\mathcal{V}'}_{\dot\beta}(w)(\tau_c)^{\dot\beta}_{~\dot\alpha}
\,}{z-w} + \cdots~~,
\\
\label{ope1+}
&J^{(+)}_3(z) \,{\mathcal{V}'}^{\dot\alpha}(w) = -\,\ii\,\frac{
\epsilon\,{\mathcal{V}'}^{\dot\alpha}(w)}{z-w} + \cdots~~,
\end{align}
\end{subequations}
which show that the new vertices form again a doublet of
$\mathrm{SU}(2)_-$ but carry opposite $\mathrm{U}(1)_+$ charge
with respect to the old vertices $\mathcal{V}$.

In complete analogy with what we did on the 3/3 strings, and in
order to be consistent with that choice, also here we treat the
two types of boundary conditions for the ($-1$)/3
strings in a symmetric way, and thus consider the following projected
vertex operator
\begin{equation}
\label{vertexw1}
V_{w} = \frac{g_0}{\sqrt{2}}\,w_{\dot\alpha}\,
\frac{\mathcal{V}^{\dot\alpha} +
{\mathcal{V}'}^{\dot\alpha}}{\sqrt{2}}~~,
\end{equation}
where we have inserted a polarization $w_{\dot\alpha}$ and a
normalization $g_0/\sqrt{2}$. In this case, however, differently from what we did
for the vertices of the 3/3 strings, we cannot simply identify
${\mathcal{V}}$ and ${\mathcal{V}'}$,
since they have different quantum numbers.
As we will see in the following sections, the
vertex (\ref{vertexw1}) correctly describes the moduli $w_{\dot\alpha}$ of the ADHM
construction for the non-commutative gauge theory, and represents
the generalization of the vertex (\ref{vertexw}) when a
self-dual $B_{\mu\nu}$ background is present.

A few remarks are in order at this point.
The projected vertex $\big(\mathcal{V}
+ {\mathcal{V}'}\big)$ is a doublet of
$\mathrm{SU}(2)_-$, but it clearly does not have a definite $\mathrm{U}(1)_+$
charge. Notice however that in any disk amplitude such a vertex must always
be accompanied by its conjugate for consistency of the Chan-Paton structure,
and hence
all relevant quantities of the ADHM construction involving $w_{\dot\alpha}$
(like constraints, explicit expression of the instanton solution, ...) are actually sensible
only to the \emph{expectation value} of $J_3^{(+)}$ between projected states,
which indeed vanishes.
In other words the polarization appearing in (\ref{vertexw1}) has effectively
the correct quantum numbers of the ADHM moduli $w_{\dot\alpha}$.
We will see an explicit example of this fact in
section \ref{secn:solution}.

This analysis can be easily repeated in the R sector of the mixed
string. Here one finds a physical GSO projected and symmetrized vertex
given by
\begin{equation}
\label{vertexmu1}
V_{\mu} = \frac{g_0}{\sqrt{2}}\,\mu^A\,
\frac{\mathcal{V}_A +
\mathcal{V}'_A}{\sqrt{2}}~~,
\end{equation}
where $\mathcal{V}_A$ is defined in (\ref{vmu}) and
$\mathcal{V}'_A$ is its analogue with the {\bf N(b)} boundary
conditions. The vertex (\ref{vertexmu1}) correctly describes the fermionic ADHM
moduli $\mu^A$ in the non-commutative gauge theory.

We conclude by mentioning that the 3/($-1$) strings can be treated
in the same way with a simple exchange of the boundary conditions at
$\sigma=0$ and $\sigma=\pi$, and that the ADHM moduli $\overline
w_{\dot\alpha}$ and $\overline \mu^A$ are described by the conjugates
of the vertices (\ref{vertexw1}) and
(\ref{vertexmu1}).

Finally, if the $B_{\mu\nu}$ background is anti-self-dual and hence
the Lorentz group is broken to $\mathrm{SU}(2)_+ \times
\mathrm{U}(1)_-$, the physical vertex operators of the NS sector
turn out to be doublets of $\mathrm{SU}(2)_+$ with charge under $\mathrm{U}(1)_-$
and can be used to describe the ADHM moduli $w_\alpha$ and
$\overline w_\alpha$ of non-commutative anti-instantons. The
transformation properties and charges of the physical NS vertices
are summarized in Table \ref{tab:2}.

\TABLE{\centerline{
\label{tab:2}
\begin{tabular}{|c|c|c|}
\hline
 &$\mathrm{U}(1)_+$&
 $\mathrm{SU}(2)_-$\\
\hline
${\mathcal V}^{\dot \alpha}$&$\epsilon$ & $\mathbf{2}$ \\
${\mathcal{V}'}^{\dot\alpha}$&$-\epsilon$ & $\mathbf{2}$ \\
\hline
\hline
&$\mathrm{SU}(2)_+$&$\mathrm{U}(1)_-$
 \\
\hline
${\mathcal V}^{\alpha}$&$\mathbf{2}$ &$\epsilon$  \\
${\mathcal{V}'}^{\alpha}$&$\mathbf{2}$ & $-\epsilon$ \\
\hline
\end{tabular}
\caption{Transformation properties of the physical NS vertices under the unbroken part of
the Lorentz group in a self-dual ($\epsilon_1=\epsilon_2=\epsilon$)
or anti-self-dual ($\epsilon_1=-\epsilon_2=\epsilon$) background.}
}}

\vskip 0.6cm
\subsubsection{The issue of stability}
\label{subsec:stability}

The previous discussion shows that in a self-dual $B$ background
the NS sector of the mixed strings contains only the moduli
$w_{\dot\alpha}$ and $\overline w_{\dot\alpha}$ associated to an instanton, while in
an anti-self-dual background it contains only the moduli
$w_{\alpha}$ and $\overline w_{\alpha}$ associated to an
anti-instanton. In other words, the explicit string realization of
the ADHM construction seems to be
applicable only to configurations where the non-commutative gauge field strength
and the $B$ background have the \emph{same} duality properties.

We now explain the origin and the physical meaning of this fact. Let us first recall that
instantons and anti-instantons are realized respectively by
systems of D3/D($-1$) branes and systems of D3/anti-D($-1$)
branes, and that the corresponding mixed strings are characterized by
a different GSO projection. Indeed,
\begin{equation}
{\cal P}_{\rm GSO}=\frac{1\pm(-1)^F}{2}
\label{GSO}
\end{equation}
where the $+$ sign applies to the D3/D($-1$) system, and
the $-$ sign to the D3/anti-D($-1$) system. It is therefore
obvious that if a state survives the GSO projection of the
D3/D($-1$) system, this same state is removed by the GSO projection
of the D3/anti-D($-1$) system, and vice-versa.
Let us consider the NS sector and fix our conventions in such a way that
\begin{equation}
(-1)^F\,\ket{\epsilon_1,\epsilon_2}_{\NS} =
-\,\ket{\epsilon_1,\epsilon_2}_{\NS}
\label{f-parity}
\end{equation}
for $\epsilon_1,\epsilon_2>0$. With this choice, the excited states
$\overline\Psi_{\dot 1\,,\,-\epsilon_1}\ket{\epsilon_1;\epsilon_2}_{\rm
NS}$ and $\overline\Psi_{\dot 2\,,\,-\epsilon_2}\ket{\epsilon_1;\epsilon_2}_{\rm
NS}$, which are physical for $\epsilon_1=\epsilon_2$, are selected
by the GSO projection of the D3/D($-1)$ branes.
If we follow $\ket{\epsilon_1,\epsilon_2}_\NS$ (which is
annihilated by all positive modes and in particular by
$\Psi^{\dot 2}_{\epsilon_2}$)
when, say, $\epsilon_2$ decreases and becomes negative, we
find that it is mapped to the state
$\Psi^{\dot 2}_{\epsilon_2}\ket{\epsilon_1,\epsilon_2}_\NS$;
thus, the very same definition of $(-1)^F$ we used in (\ref{f-parity}),
leads to
\begin{equation}
(-1)^F\,\ket{\epsilon_1,\epsilon_2}_{\NS} =
+\,\ket{\epsilon_1,\epsilon_2}_{\NS}
\label{f-parity1}
\end{equation}
for $\epsilon_1>0$ and $\epsilon_2<0$. If we now let $\epsilon_1$
become negative as well, the $F$ parity of the vacuum returns to be $(-1)$
as in (\ref{f-parity}). We can then conclude that if $\epsilon_1$ and
$\epsilon_2$ have the same sign, the vacuum $\ket{\epsilon_1,\epsilon_2}_\NS$
has $F$-parity $(-1)$, whereas if $\epsilon_1$ and
$\epsilon_2$ have different signs, the vacuum $\ket{\epsilon_1,\epsilon_2}_\NS$
has $F$-parity $(+1)$. Thus, in a self-dual background the
physical states, which are forced to stay at the first excited level of the NS
sector,
survive the GSO projection appropriate for instantons while in an
anti-self-dual background they survive the GSO projection of
anti-instantons.

This asymmetry has a deep physical meaning: indeed, it is related
to the fact that D3/D($-1$) systems are stable only in a
self-dual background, while D3/anti-D($-1$) systems are stable
only in an anti-self-dual background~\footnote{A discussion on
related issues appears also in Ref.\cite{Seiberg:1999vs}.}.
To investigate the stability of these systems,
we compute the one-loop free energy for the oriented open strings stretching between
the D3 branes and the (anti-)D-instantons in presence of a $B$
background. This free energy is given by
\begin{equation}
\label{fe1}
\mathcal{F} = -\int_0^{\ii\infty}
\frac{d\tau}{2\tau} \,
\Big[\Tr_{\NS}~ q^{L_0 - \frac{c}{24}} \pm \Tr_\NS~(-1)^F q^{L_0 -
\frac{c}{24}} -
\Tr_{\R}~ q^{L_0 - \frac{c}{24}}\Big]
\end{equation}
where $q={\rm e}^{2\pi\ii\tau}$ and $c$ is the total central
charge of the CFT in the light-cone.
Notice that in (\ref{fe1}) we have not written the trace with $(-1)^F$ in the R sector since it
vanishes due to the fermionic zero-modes in the six transverse
directions. According to (\ref{GSO}), the $+$ sign in $\mathcal{F}$  refers to the
D3/D($-1$) case, while the $-$ sign refers to the D3/anti-D($-1$) case.

Let us discuss the contribution to the traces from the
four mixed directions which feel the effect of the $B$
background. Actually, we can focus just on the fermionic piece,
since all the bosonic contributions to the integrand of
(\ref{fe1}) are common to the various sectors.
In computing the traces for the CFT of the twisted fermions
we have to take into account in a symmetric way the two
types of boundary conditions, {\bf{N(a)}} and {\bf{N(b)}}, which they can satisfy
(see Eqs. (\ref{Neu fermionA})
and (\ref{Neu fermionB})). In practice, this means that
\begin{equation}
\label{av_AB}
\Tr_{\rm NS, R} \to \frac{\Tr_{\rm NS, R}^{\bf{(a)}} +
\Tr_{\rm NS, R}^{\bf{(b)}} }{2}~~.
\end{equation}
Using (\ref{L0ferm}) for the two complex fermions $\Psi^1$ and $\Psi^2$ and introducing
the standard Jacobi $\theta$-functions, for any value of $\epsilon_1$ and
$\epsilon_2$ we find
\begin{subequations}
\label{cbPsi}
\begin{align}
&\Tr_{\rm NS}^{\bf(a)}\,\,q^{L_0 - \frac{c}{24}}
=
q^{\frac{\epsilon_1^2+\epsilon_2^2}{2}}~
\frac{\theta_2\left(\epsilon_1\tau|\tau\right)\theta_2\left(\epsilon_2\tau|\tau\right)}
{\eta(\tau)^2}~~,
\label{trNS}
\\
&\Tr_{\rm NS}^{\bf(a)}\,\,(-1)^F\,q^{L_0 - \frac{c}{24}}
=
q^{\frac{\epsilon_1^2+\epsilon_2^2}{2}}~
\frac{\theta_1\left(\epsilon_1\tau|\tau\right)\theta_1\left(\epsilon_2\tau|\tau\right)}
{\eta(\tau)^2}~~,
\label{trNSF}
\\
&\Tr_{\rm R}^{\bf(a)}\,\,q^{L_0 - \frac{c}{24}} =
q^{\frac{\epsilon_1^2+\epsilon_2^2}{2}}~
\frac{\theta_3\left(\epsilon_1\tau|\tau\right)\theta_3\left(\epsilon_2\tau|\tau\right)}
{\eta(\tau)^2}
\label{trR}
\end{align}
\end{subequations}
where $\eta$ is the Dedekind's function.
The partition functions for the boundary conditions of type {\bf{N(b)}} can
be simply obtained from (\ref{cbPsi}) by reversing the signs of
both $\epsilon_1$ and $\epsilon_2$. Taking into account the
parity properties of the $\theta$-functions, it is immediate to
show that $\Tr^{\bf(b)}=\Tr^{\bf(a)}$ in all sectors.

Using (\ref{cbPsi}) and the standard results for the four transverse directions,
in the case of instantons, {\it i.e.} for D3/D($-1$) systems,
we find in the end that the integrand of (\ref{fe1}) is proportional to
\begin{equation}
\begin{aligned}
\label{fe2}
& \theta_2(\epsilon_{1}\tau|\tau)\theta_2(\epsilon_{
2}\tau|\tau) \left[\theta_3(0|\tau)\right]^2
\pm \theta_1(\epsilon_{1}\tau|\tau)\theta_1(\epsilon_{2}\tau|\tau)
\left[\theta_4(0|\tau)\right]^2
\\
& - \theta_3(\epsilon_{1}\tau|\tau)\theta_3(\epsilon_{
2}\tau|\tau)\left[\theta_2(0|\tau)\right]^2~~.
\end{aligned}
\end{equation}
In the case of D$(-1)$ branes, {\it i.e.} with the upper sign in
(\ref{fe2}), this expression identically can vanish only for a self-dual background
($\epsilon_{1} =
\epsilon_{2}=\epsilon$) thanks to the identity
\begin{equation}
\label{theta_id}
\left[\theta_2(\epsilon\tau|\tau)\right]^2
\left[\theta_3(0|\tau)\right]^2
+\left[\theta_1(\epsilon\tau|\tau)\right]^2
\left[\theta_4(0|\tau)\right]^2
- \left[\theta_3(\epsilon\tau|\tau)\right]^2
\left[\theta_2(0|\tau)\right]^2 = 0~~.
\end{equation}
If we consider instead anti-D($-1$) branes, {\it i.e.} if we take
the lower sign in (\ref{fe2}), we see that
the interaction energy vanishes only for an {anti-self-dual} background
$\epsilon_{1} = -\epsilon_{2} =
\epsilon$, since $\theta_1$ is odd in its first argument while
$\theta_2$ and $\theta_3$ are even.
In conclusion we see that there must be a precise relation between
the charge of the D-instantons and the (anti-)self-duality of the $B$
background in order to have a stable brane system.

\vskip 0.8cm
\section{Non-commutative gauge instantons from open strings}
\label{secn:solution}
In this section we are going to present an explicit realization of
instantons in non-commutative gauge theories using the open
strings of the D3/D($-1$) systems described above, thus
extending the analysis of Ref. \cite{Billo:2002hm} to branes in a
background $B$ field. For definiteness, we discuss in detail the stable
case of a D3/D($-1$) system in a self-dual background, starting
from the ADHM constraints.

\vskip 0.6cm
\subsection{The ADHM constraints}
\label{subsec:mod_ADHM}
The ADHM measure on the instanton moduli space can be derived from
scattering amplitudes involving all excitations of the open strings with at least one
end point on the D-instantons. As mentioned in section \ref{subsec:11},
the $B$ background does not have any effect on the $(-1)$/$(-1)$ strings
and thus their contribution to the ADHM measure is
identical to that of the undeformed (commutative) theory.
On the other hand, the mixed ($-1$)/3 or 3/($-1$) sectors feel the
presence of the $B$ field, and thus to find the non-commutative modifications
in the ADHM measure we have to compute only the amplitudes that involve mixed moduli.
For example, let us consider the
coupling among $w_{\dot\alpha}$, ${\overline w}_{\dot\alpha}$ and the auxiliary fields
$D_c^{-}$, which play the r\^ole of a Lagrange
multipliers for the bosonic ADHM constraints. The vertex operators
for $w_{\dot\alpha}$ and ${\overline w}_{\dot\alpha}$ are given
in (\ref{vertexw1}) and its conjugate, while vertex for $D_c^{-}$ is
the same as the undeformed one given in (\ref{vertDmod}). However, in
the self-dual background it is more convenient to rewrite the latter in the following manner
\begin{equation}
\label{vertDnew}
V_D = -2\, D^-_c \,J^{(-)}_c
\end{equation}
using the $\mathrm{SU}(2)_-$ currents defined in
(\ref{ferm_curr}).
The coupling among the moduli we are considering is explicitly given by
\begin{equation}
\label{ampliwDwb}
\lvev V_{w} V_D V_{\overline w}\rvev \equiv
C_0 \int \frac{dy_1 dy_2 dy_3}{dV_{\mathrm{CKG}}}
~\tr \langle V_{w}(y_1) V_D(y_2) V_{\overline w}(y_3)
\rangle~~,
\end{equation}
where $C_0 = 2/g_0^2$ is the normalization of the mixed disk amplitudes
in our present conventions
(see {\it e.g.} Ref. \cite{Billo:2002hm} for further details).
Moreover, $dV_{\rm CKG}$ is the $\mathrm{SL}(2,\mathbb{R})$ invariant volume
element, and the trace is over the $\mathrm{U}(N)$ Chan-Paton factors.
Using the expression for the various vertex operators, after a few
straightforward steps, the amplitude (\ref{ampliwDwb}) becomes
\begin{equation}
\label{ampliwDwb2}
\begin{aligned}
\lvev V_{w} V_D V_{\overline{w}}\rvev & = -\,
D_c^-\,\tr\big({w}_{\dot\alpha}\,\overline w_{\dot\beta}\big)\,
(y_1 - y_2)(y_2 - y_3)(y_3-y_1)\\
& ~~~\times
\Big[
\langle {\cal V}^{\dot\alpha}(y_1)\,J^{(-)}_c(y_2)
\,\overline{\cal V}^{\dot\beta}(y_3)
\rangle~
\,+\,
\langle {\cal V'}^{\dot\alpha}(y_1)\,J^{(-)}_c(y_2)
\,\overline{\cal V'}^{\dot\beta}(y_3)
\rangle
\Big]~~.
\end{aligned}
\end{equation}
As discussed in section \ref{subsubsec:asdbkg}, the vertices ${\cal V}$
and ${\cal V'}$ transform in the same way under
the currents $J^{(-)}_c$, and indeed from (\ref{opeJ-}) and
(\ref{ope1-}) one can show that
\begin{equation}
\label{wjw}
\begin{aligned}
\langle {\cal V}^{\dot\alpha}(y_1)\,J^{(-)}_c(y_2)
\,\overline{\cal V}^{\dot\beta}(y_3)
\rangle&=
\langle {\cal V'}^{\dot\alpha}(y_1)\,J^{(-)}_c(y_2)
\,\overline{\cal V'}^{\dot\beta}(y_3)
\rangle\\
&=\frac{\ii}{2}\, \frac{(\tau^c)^{\dot\alpha\dot\beta}}
{(y_1-y_2)\,(y_1-y_3)\,(y_2-y_3)}~~.
\end{aligned}
\end{equation}
Inserting this result into (\ref{ampliwDwb2}), we simply obtain
\begin{equation}
\label{wDwbres}
\lvev V_{\bar w} V_D V_{w}\rvev =
\ii \, D_c^- \tr\left(w_{\dot\alpha}
(\tau^c)^{\dot\alpha}_{\phantom{\dot\alpha}\dot\beta}
\overline{w}^{\dot\beta}\right) =
\ii D_c^- W^c~,
\end{equation}
which is exactly the same amplitude of the undeformed theory
\cite{Billo:2002hm}.

With similar explicit calculations one can check that all other
disk diagrams involving mixed moduli are not affected by the self-dual
background, and thus the complete non-commutative ADHM moduli action is the same
as in the commuting theory.
In particular, the bosonic and fermionic ADHM constraints are
still given by (\ref{bosADHM}) and (\ref{fermADHM}), respectively.
As we shall see in the next section, this string result is in agreement with the explicit ADHM
construction for non-commutative instantons \cite{Chu:2001cx}.

\vskip 0.6cm
\subsection{The instanton profile in a self-dual background}
\label{subsec:profile}
In string theory the classical instanton solutions are obtained from
1-point diagrams that describe the emission of the gauge
fields from mixed disks
\cite{Billo:2002hm}. The simplest
example of such a mixed disk diagram is represented in Fig.
\ref{fig:md}, which describes the emission of a gauge vector.

\FIGURE{
\centerline{
\psfrag{alabel}{\emph{a)}}
\psfrag{blabel}{\emph{b)}}
\psfrag{a}{\small $\hat a$}
\psfrag{mu}{\small $\mu$}
\psfrag{w}{\small $\bar w$}
\psfrag{wb}{\small $w$}
\psfrag{p}{\small $p$}
\includegraphics[width=0.25\textwidth]{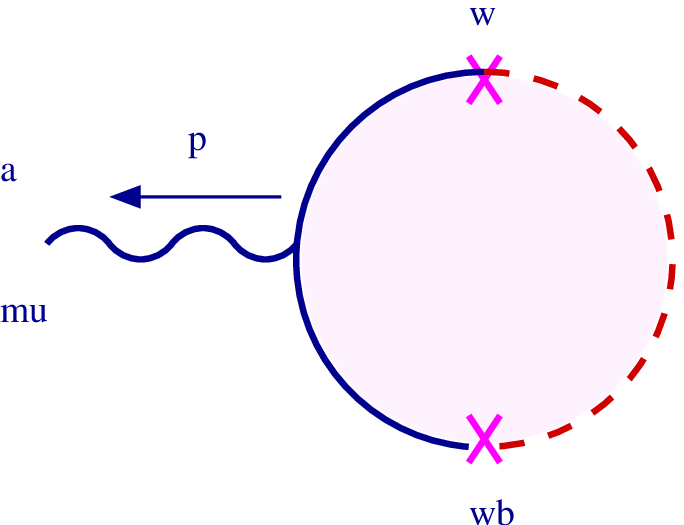}
\caption{\small The mixed disk diagram that describes the emission of a
gauge vector field $A_\mu^{\hat a}$ with momentum $p$ represented by the
outgoing wavy line.
}
\label{fig:md}
}}

For simplicity, in the following we discuss only
instantons with charge $k=1$ in the non-commutative
$\mathrm{U}(2)$ gauge theory, but our analysis can be
extended to the general case without any problems.
The amplitude described in Fig. \ref{fig:md}
explicitly reads
\begin{equation}
\label{dia1}
A^{\hat a}_\mu(p) =
\lvev V_{\overline w}\,{\cal V}^{(0)}_{A^{\hat a}_\mu}(-p)\,V_{w}
\rvev
\end{equation}
where ${\cal V}^{(0)}_{A^{\hat a}_\mu}(-p)$ is the gluon vertex operator in the 0-superghost
picture with {outgoing} momentum and {without} polarization, {\it
i.e.}
\begin{equation}
\label{vert2}
{\cal V}^\ppz_{A^{\hat a}_\mu}(-p) = 2\ii\, T^{\hat a}\left(
\partial X_\mu \,-\, \ii \,p\cdot \psi\, \psi_\mu\right)\,
\ee^{-\ii p\cdot X}~~,
\end{equation}
where $T^{\hat a}$ is the adjoint $\mathrm{U}(2)$ Chan-Paton
factor. With the insertion of such a vertex, the disk amplitude (\ref{dia1})
carries the Lorentz structure and the quantum numbers that are
appropriate for an emitted gauge vector field.
The correlation function in (\ref{dia1}) receives contribution only from
the $\psi^\mu\psi^\nu$ part of (\ref{vert2}), which
again can be conveniently rewritten in terms of the $\mathrm{SU}(2)_\pm$
fermionic currents (\ref{ferm_curr}). Thus, the relevant part of the gluon vertex is
\begin{equation}
\label{vert21}
{\cal V}^\ppz_{A^{\hat a}_\mu}(-p) \sim 2p^{\nu}\, T^{\hat a}
\left(
\eta_{\nu\mu}^c J_c^{(+)}\,+\,\bar\eta_{\nu\mu}^c J_c^{(-)}
\right)
\ee^{-\ii p\cdot X}~~.
\end{equation}
Notice, that differently from the auxiliary vertex
(\ref{vertDnew}), the gluon vertex (\ref{vert21}) depends
both on the $\mathrm{SU}(2)_+$ and on the $\mathrm{SU}(2)_-$ currents. This fact has
important consequences, as we shall see momentarily.

The calculation of the $J_c^{(-)}$ contribution to the amplitude
(\ref{dia1}) essentially coincides with the one outlined in the
previous subsection for the amplitude (\ref{ampliwDwb2}) (see also
Ref. \cite{Billo:2002hm} for further details). Since
the vertices ${\cal V}$ and ${\cal V'}$ appearing in $V_w$ and $V_{\overline w}$
behave in the same way
under $\mathrm{SU}(2)_-$, they produce an identical contribution to the gluon emission
which is given by
\begin{equation}
\label{corr5}
{\cal A}^{\hat a\,(-)}_\mu(p) = \frac{\ii}{2}\, (T^{\hat a})^{v}_{~u}\,p^\nu
\, \bar\eta^c_{\nu\mu}
\left(w_{\dot\alpha}^{~u}\,(\tau_c)^{\dot\alpha}_{~\dot\beta}\,
\bar w^{\dot \beta}_{~v}\right)~~.
\end{equation}
Imposing the bosonic ADHM constraints (\ref{bosADHM}) on $w$ and $\overline w$,
we can show that the matrices
\begin{equation}
\label{gf2}
(T_c)^{u}_{~v} \equiv \frac{1}{2 \rho^2} \left(w_{\dot\alpha}^{~u}\,
(\tau_c)^{\dot\alpha}_{~\dot\beta}
\,\bar w^{\dot \beta }_{~v}\right)~~,
\end{equation}
where $\rho^2 \equiv (\bar w^{\dot\alpha}_{~u}\, w_{\dot\alpha}^{~u})/2$,
satisfy the $\mathrm{SU}(2)$ algebra, so that (\ref{corr5}) can be rewritten as
\begin{equation}
\label{gf3}
{\cal A}^{\hat a\,(-)}_\mu(p) = \ii\rho^2\,\Tr\, (T^{\hat a}\, T_c) \,
p^\nu\,\bar\eta^c_{\nu\mu}~~.
\end{equation}
Decomposing the adjoint $\mathrm{U}(2)$
index $\hat a=(0,a)$ into its $\mathrm{U}(1)$ and $\mathrm{SU}(2)$
parts, we see that only the $\mathrm{SU}(2)$
components are non-vanishing,
namely
\begin{equation}
\label{gf6}
{\cal A}^{a\,(-)}_\mu(p) = -\frac{\ii}{2}\,\rho^2\,
\bar\eta^a_{\mu\nu}p^\nu~~~~~,~~~~~
{\cal A}^{0\,(-)}_\mu(p) = 0~~.
\end{equation}

Things are different instead for the $J_c^{(+)}$ contributions to
the amplitude (\ref{dia1}): this is precisely the point where the non-trivial
structure of the vertices $V_w$ and $V_{\overline w}$ given in (\ref{vertexw1})
shows its relevance. In fact, while the correlators of ${\cal V}$ and ${\cal V}'$
with the step operators $J^{(+)}_{\pm}$ are vanishing, {\it i.e.}
\begin{equation}
\label{jpm}
\langle
\overline{\cal V}_{\dot\beta}(y_1)\,J^{(+)}_{\pm}(y_2){\cal V}^{\dot\alpha}(y_3)
\rangle ~=~
\langle
\overline{\cal V'}_{\dot\beta}(y_1)\,J^{(+)}_{\pm}(y_2){\cal V'}^{\dot\alpha}(y_3)
\rangle =0~~,
\end{equation}
their correlators with $J^{(+)}_{3}$ are instead non-trivial, as one can see
from the OPE's (\ref{chargew}) and (\ref{ope1+}). Indeed we have
\begin{equation}
\label{j3}
\begin{aligned}
\langle
\overline{\cal V}_{\dot\beta}(y_1)\,J^{(+)}_{3}(y_2){\cal V}^{\dot\alpha}(y_3)
\rangle &=-\,\langle
\overline{\cal V'}_{\dot\alpha}(y_1)\,J^{(+)}_{3}(y_2){\cal V'}^{\dot\alpha}(y_3)
\rangle \\
&=\,\frac{\ii\,\epsilon\,\delta^{\dot\alpha}_{\dot\beta}}
{(y_1-y_2)(y_1-y_3)(y_2-y_3)}~~.
\end{aligned}
\end{equation}
Thus, the $J^{(+)}_c$ part of the gluon emission is
\begin{equation}
\label{a+}
{\cal A}^{{\hat a}\,(+)}_\mu(p) = \pm\,\frac{\ii\,\epsilon}2 \, p^\nu \eta^3_{\mu\nu}
(T^{\hat a})^{v}_{~u}\bar w^{\dot\alpha\,u}\, w_{\dot\alpha\,v}
\end{equation}
where the $+$ and $-$ signs apply to the contributions of the
${\cal V}$ and ${\cal V}'$ vertices, respectively.
If we impose the bosonic ADHM constraints, we can show that the matrix
$\bar w^{\dot\alpha\,u}\, w_{\dot\alpha\,v}$ is proportional to the
identity, {\it i.e.} $\bar w^{\dot\alpha\,u}\,
w_{\dot\alpha\,v}\,=\,\rho^2\,\delta^u_v$, and
so only the $\mathrm{U}(1)$ part of (\ref{a+}) is non-vanishing, namely
\begin{equation}
\label{a01}
{\cal A}^{0\,(+)}_\mu(p) = \pm\,\ii\,\epsilon \, p^\nu \eta^3_{\mu\nu}\rho^2
~~~~~,~~~~~{\cal A}^{a\,(+)}_\mu(p)=0~~.
\end{equation}
While the result (\ref{gf6}) is somehow expected (and in
agreement with field theory calculations), the presence of a
$\mathrm{U}(1)$ component of the form (\ref{a01}) is puzzling
since one does not expect a non-vanishing abelian part
in the gluon emission at this order. However, in our string realization of the
non-commutative ADHM instantons the vertices
${\cal V}$ and ${\cal V}'$, which correspond to the two types of
Neumann boundary conditions in the presence of a $B$ field, are
treated in a symmetric way and their respective contributions
to a given amplitude must be added together. Thus, the complete
$J^{(-)}$ piece of the gluon emission is simply twice the
result given in (\ref{gf6}) for each sector, while the total
$J^{(+)}$ contribution vanishes because the ${\cal V}$ and ${\cal V}'$
parts exactly cancel each other. In other words the full amplitude
(\ref{dia1}) is
\begin{equation}
{A}^{a}_\mu(p) = -\,{\ii}\,\rho^2\,
\bar\eta^a_{\mu\nu}p^\nu~~~~~,~~~~~
{A}^{0}_\mu(p) = 0~~.
\label{totalampl}
\end{equation}

As explained in Ref. \cite{Billo:2002hm}, to obtain the space-time
profile of the instanton we must take the Fourier transform of the
momentum space amplitude after inserting a gluon propagator
$\delta_{\mu\nu}/p^2$; in this way we get
\begin{equation}
\label{gf1}
A^{a}_\mu(x) = \int \frac{d^4 p}{(2\pi)^2} \,
A^{a}_\mu(p) \,\frac{1}{p^2}\,\ee^{\ii p\cdot x}
= 2\rho^2\, \bar\eta^a_{\mu\nu} \,
\frac{x^\nu}{|x|^4}~~~~~,~~~~~
{A}^{0}_\mu(x) = 0~~.
\end{equation}
In the following section we will see that (\ref{gf1}) represents
the leading term in the large distance expansion ($|x|\gg\rho$) of
the classical  solution in the singular gauge for an instanton
of size $\rho$ and charge $k=1$ in
the non-commutative $\mathrm{U}(2)$ theory. The fact that the instanton field is
in the singular gauge is not surprising since in our D-brane
set-up gauge instantons arise from D-instantons which are point-like objects
localized inside the world-volume of the D3 branes~\cite{Billo:2002hm}.

Notice that the gauge field in (\ref{gf1}) does not depend on the
non-commutativity parameter $\theta$ and is the same as the leading term at large distance
of the BPST instanton of the ordinary $\mathrm{SU}(2)$
Yang-Mills theory in the singular gauge
\cite{Belavin:fg}. However, the presence of a non-trivial $B$
background is not irrelevant and it shows up in the sub-leading
terms of the instanton solution.
Indeed, higher order contributions in the
large distance expansion of the instanton profile can be obtained by
sewing the leading source term with the vertices of
the non-commutative gauge theory, as indicated for example in Fig.
\ref{fig:2ndorder}. There, two gauge vector fields emitted from two disks
recombine through the non-commutative 3-gluon vertex and yield the
second order correction to the instanton profile.
Since the non-commutative vertex contains a part proportional to the
$d_{\hat a \hat b \hat c}$-symbols of the gauge group,
a gluon can be emitted at second order also along the $\mathrm{U}(1)$
direction, even if at the first order only the non-abelian components
were produced. 

\FIGURE{
\centerline{
\psfrag{alabel}{\emph{a)}}
\psfrag{blabel}{\emph{b)}}
\psfrag{a}{\scriptsize $a$}
\psfrag{b}{\scriptsize $b$}
\psfrag{c}{\scriptsize $c$}
\psfrag{0}{\scriptsize $0$}
\psfrag{pmom}{\scriptsize $p$}
\psfrag{qmom}{\scriptsize $p-q$}
\psfrag{p-qmom}{\scriptsize $q$}
\psfrag{mu}{\scriptsize $\mu$}
\psfrag{nu}{\scriptsize $\nu$}
\psfrag{rho}{\scriptsize $\rho$}
\psfrag{w}{\scriptsize $\bar w$}
\psfrag{wb}{\scriptsize $w$}
\psfrag{cos}{\scriptsize $\cos p\wedge q$}
\psfrag{sin}{\scriptsize $\sin p\wedge q$}
\includegraphics[width=0.70\textwidth]{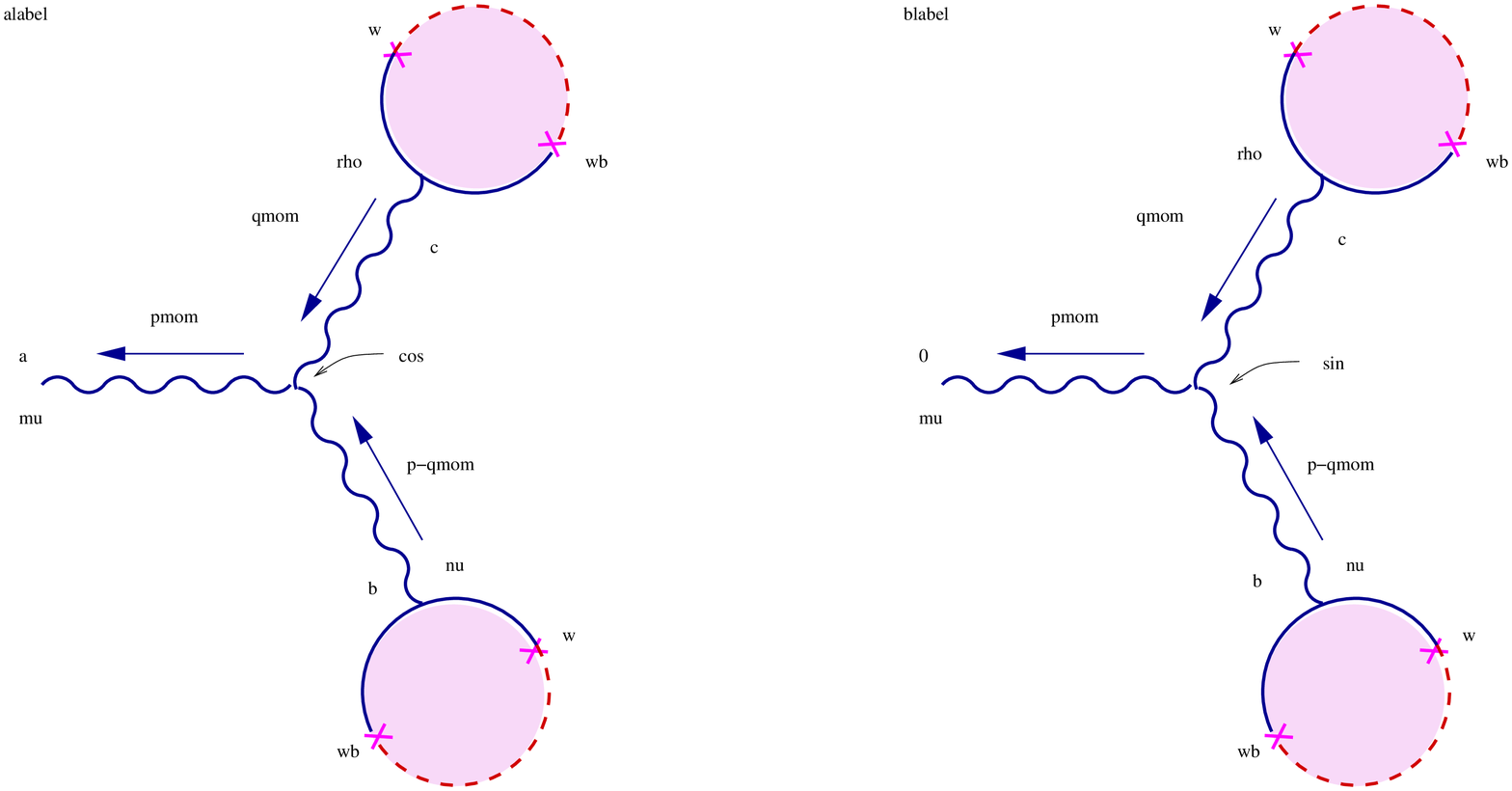}
\caption{\small The diagrams that account for
sub-leading corrections in the large-distance expansion of the
gluon emission amplitude. The diagram \emph{a)} refers to the
emission of the $\mathrm{SU}(N)$ gluon, while the diagram \emph{b)} the
$\mathrm{U}(1)$ part. The 3-gluon vertices to be used, here only symbolically indicated,
are more explicitly described in Fig. \ref{fig:ncv}.}
\label{fig:2ndorder}
}}

Let us consider in particular the diagram \emph{a)} in which
the two disks are sewn together with the part of the 3-gluon vertex
that is proportional to the structure constants of the gauge group (see Fig. \ref{fig:ncv}).
This diagram yields a second order correction to the $\mathrm{SU}(2)$ field given by
\begin{equation}
\label{2o1}
\begin{aligned}
{A^a_\mu (p)}^{(2)} &
= \frac{\ii}{2}
\int \!\!\frac{d^4 q}{(2\pi)^2}
\epsilon^{abc}\cos\left(p\!\wedge\! q\right)
\Bigl[(p \!-\! 2q)_\mu\delta_{\nu\rho} + (p \!+ \! q)_\rho\delta_{\mu\nu}
 + (q \! - \! 2 p)_\nu\delta_{\rho\mu}\Bigr]
\\
& ~~~~~~\times
\frac{1}{q^2}\,\,{A^b_\nu (q)}^{(1)}\,
\frac{1}{(p-q)^2}\,\,{A^c_\rho (p-q)}^{(1)}~~,
\end{aligned}
\end{equation}
where we have included a symmetry factor of $1/2$ and
denoted with a superscript ${}^{(1)}$ the first-order fields of
(\ref{gf1}), and by ${}^{(2)}$ the second-order correction we are computing.
Using (\ref{gf1}), after some standard algebra, we find
\begin{equation}
\label{2o2}
{A^a_\mu (p)}^{(2)} 
= - \ii \rho^4\overline\eta^a_{\mu\nu} \int \!\!\frac{d^4 q}{(2\pi)^2}
\frac{\cos\left(p\!\wedge\! q\right)}{q^2(p-q)^2}
\,\,\Bigl[(2 p\cdot q - p^2) \,q^\nu + (p\cdot q - 2 q^2)
\,p^\nu\Bigr]~~.
\end{equation}
The integral over $q$ can be computed in dimensional regularization upon expanding
\begin{equation}
\label{expcos}
\cos\left(p\!\wedge\! q\right) = 1 - \frac 12 \Big(\frac 12 p_\mu
\theta^{\mu\nu} q_\nu\Big)^2 + \ldots~,
\end{equation}
where $\theta$ is the non-commutativity parameter (\ref{theta}), and the result
in $d$ dimensions is
\begin{equation*}
\label{dpres}
{A^a_\mu (p)}^{(2)}
= \ii \rho^4\overline\eta^a_{\mu\nu} p^\nu  {(p^2)}^{\frac d2 -1}
\sum^\infty_{m=0}\frac{{(-|p|^4)}^{m}}{m!}
\frac{\pi\,\Gamma\left(m+\frac d2\right)}{{2^{\frac
d2}}\sin\left(\pi(m+\frac{d}{2}) \right)\Gamma(2m+d-1)}{\left(\frac{\theta^2}{16}\right)}^m
~~.
\end{equation*}
Taking the Fourier transform after inserting the gluon propagator,
we find
\begin{equation}
\label{dxres}
\begin{aligned}
{A^a_\mu (x)}^{(2)}
&= \lim_{d\to 4}\int \frac{d^d p}{(2\pi)^{\frac d2}} \,
A^{c}_\mu(p)^{(2)} \,\frac{1}{p^2}\,\ee^{\ii p\cdot x}
\\&= -2 \rho^4 \overline\eta^a_{\mu\nu} \frac{x^\nu}{|x|^6}
\sum^\infty_{m=0}(m+1)(2m)!{\left(\frac{\theta}{|x|^2}\right)}^{2m}\\
&= -2 \rho^4 \overline\eta^a_{\mu\nu} \frac{x^\nu}{|x|^6} \left(1 +
\frac{4\theta^2}{|x|^4} + \ldots \right)
~~.
\end{aligned}
\end{equation}
In the first term of the last line we recognize the sub-leading
term in the large distance expansion ($|x|\gg\rho$) of the
standard $\mathrm{SU}(2)$ instanton in the singular gauge (see for
example Eq. (4.16) of Ref. \cite{Billo:2002hm}), while the second
term represents the non-commutative deformation. In the next
section we will show that this result is in agreement with the
explicit non-commutative ADHM instanton solution.
Notice, however, that the series (\ref{dxres}) is divergent and
thus it must be interpreted as an asymptotic expansion for
${\theta}/{|x|^2}\to 0$. As a consequence, possible
non-perturbative terms, like ${\rm e}^{-|x|^2/\theta}$ for instance,
cannot be accounted in this approach.

Let us now consider the diagram \emph{b)} of Fig.
\ref{fig:2ndorder},
which is responsible for a $\mathrm{U}(1)$ component
in the instanton profile that is absent in
the commutative case. Using the explicit expression for the
3-gluon vertex, we find
\begin{equation}
\label{1o1}
\begin{aligned}
{A^0_\mu (p)}^{(2)} &
= \frac{\ii}{2}
\int \!\!\frac{d^4 q}{(2\pi)^2}\,
\delta^{bc}\,\sin\left(p\!\wedge\! q\right)
\Bigl[(p\!- \!2q)_\mu\,\delta_\nu^\rho + (p\!+\! q)^\rho\,\delta_{\mu\nu}
 + (q \!-\! 2 p)_\nu \,\delta^{\rho}_\mu\Bigr]
\\
& ~~~~~~\times
\frac{1}{q^2}\,\,{A^b_\nu (q)}^{(1)}\,
\frac{1}{(p-q)^2}\,\,{A^c_\rho (p-q)}^{(1)}~~,
\end{aligned}
\end{equation}
which after some standard algebra becomes
\begin{equation}
\label{1o2}
{A^0_\mu (p)}^{(2)}
= \ii \rho^4 \int \!\!\frac{d^4 q}{(2\pi)^2}
\frac{\sin\left(p\!\wedge\! q\right)}{q^2(p-q)^2}
\,\,\Bigl[p_\mu (q\cdot p-3q^2) + q_\mu(6q^2-6 q\cdot p+2p^2)\Bigr]~~.
\end{equation}
Again, the integral over $q$ can be computed in dimensional
regularization after expanding $\sin\left(p\!\wedge\! q\right)$ in
powers of $\theta$. Finally, taking the Fourier transform of the resulting terms, we obtain
the following space-time dependence for the $\mathrm{U}(1)$ field
\begin{equation}
\begin{aligned}
\label{udxres2}
{A^0_\mu (x)}^{(2)}
&= -\rho^4
\frac{\theta_{\mu\nu}x^\nu}{|x|^8}\sum^\infty_{m=0}(m+1)(2m+1)!{\left(\frac{
\theta}{|x|^2}\right)}^{2m}\\
&= -\rho^4 \frac{\theta_{\mu\nu}x^\nu}{|x|^8} + \ldots ~~.
\end{aligned}
\end{equation}
In the next section we will show that this expression is in perfect agreement
with the abelian part of the
non-commutative $\mathrm{U}(2)$ ADHM instanton solution in the singular gauge
(again up to non-perturbative terms in $\theta/|x|^2$).

Our analysis can be generalized also to mixed diagrams that
contain insertions of fermionic mixed moduli and are
responsible for the emission of the other components of the gauge
vector multiplet, along the lines discussed in Ref. \cite{Billo:2002hm}.
In this way one can reconstruct the full
non-commutative superinstanton solution from string theory.

\vskip 0.8cm
\section{The ADHM construction for non-commutative instantons in the singular gauge}
\label{sec:nci_moyal}

The classical profile for non-commutative instantons can be
derived from a generalization of the standard ADHM construction
in which the non-commutative nature of space-time is properly
taken into account \cite{Chu:2001cx}. In this section we are going to briefly
present such a construction and, in order to match with the string
results we have obtained so far, we will consider specifically the non-commutative
instantons in the singular gauge.

In the standard ADHM construction of a self-dual instanton with charge $k=1$
in the ${\mathrm U}(N)$ gauge theory (see, for instance, Ref. \cite{Dorey:2002ik} for a review),
the basic object is a $(N+2)\times 2$ complex valued matrix
$\Delta(x)$ defined as
\begin{equation}
\label{defdelta}
\Delta(x)=a+ b\,x
\end{equation}
where $a$ and $b$ can always be put in the form
\begin{equation}
\label{defab}
a=\begin{pmatrix}
w\\
-a'\\
\end{pmatrix}
\qquad\qquad
b=\begin{pmatrix}
0\\
\uno
\\\end{pmatrix}
\end{equation}
in which the upper components are $N\times 2$ matrices and the lower
components are $2\times 2$ matrices. Finally, in \eqref{defdelta}
$x$ stands for the $2\times2$ matrix $x_\mu\,\sigma^\mu$ defined
in terms of the $\mathrm{SO}(4)$ spinor matrices.
The variables $w$ and $a'$ appearing in \eqref{defab} are
precisely some of the ADHM moduli for which we have presented an
explicit string realization in sections \ref{sec:d3d-1} and
\ref{sec:d3d-1wb}.
Let $\overline \Delta(x)= \Delta^\dagger(x)$ and denote by $U(x)$
a null vector of $\overline \Delta(x)$,
{\it i.e.} a solution of $\overline \Delta(x)\, U(x)=0$. Then, if the following
{completeness} and {factorization} constraints
\begin{subequations}
\begin{align}
\label{ADHM2}
&\Delta(x)\,f(x) \,\overline\Delta(x)
\,+\,
U(x)\,\overline U(x)= \uno
\\
\label{ADHM1}
&\overline\Delta(x)
\,\Delta(x)=f^{-1}(x)\uno
\end{align}
\end{subequations}
are satisfied for some (arbitrary) function $f(x)$, one can show that the gauge
potential
\begin{equation}
\label{instsolU}
A_\mu=\ii\, \overline U(x)\,\partial_\mu\,U(x)
\end{equation}
has a self-dual field strength and thus describes an instanton.

This ADHM construction can be generalized to non-commutative gauge
theories \cite{Chu:2001cx}. The only formal change in the above
equations is that now $x^\mu$ has to be interpreted as an operator
${\hat x}^\mu$, subject to the following commutation rules
\begin{equation}
\label{[x,x]}
\left[{\hat x}^\mu,{\hat x}^\nu\right]=i\theta^{\mu\nu}~,
\end{equation}
where $\theta$ is the non-commutativity parameter.
In particular this implies that \eqref{instsolU} must be replaced by
\begin{equation}
\label{instsolU2}
A_\mu=\ii\, \overline U({\hat x})\,[\hat\partial_\mu\, ,\,U({\hat x})]~~.
\end{equation}

Let us consider, as a specific example, the non-commutative theory
with gauge group $\mathrm{U}(2)$, so that we can compare this
formal construction with the explicit results obtained from string
theory in section \ref{secn:solution}.
In this case the constraint (\ref{ADHM1}) imposes a reality condition on
$a'$ and the following relation on $w$ and $\overline w$
\begin{equation}
\label{vincolo}
w_{\dot\alpha}^{\,~u}\,(\tau^c)^{\dot\alpha}_{~\dot\beta}
\, \overline w^{\,\dot\beta}_{~u}
-\overline\eta^c_{\mu\nu}\theta^{\mu\nu}=0~~.
\end{equation}
This is the non-commutative version of the bosonic ADHM constraint
\eqref{bosADHM} for $k=1$ and gauge group $\mathrm{U}(2)$.
If $\theta^{\mu\nu}$ is self-dual, the last term vanishes and
(\ref{vincolo}) reduces simply to $W^c=0$. We therefore confirm
explicitly the string results of section \ref{subsec:mod_ADHM}, where we have shown that
the ADHM constraints are not affected by a self-dual background.

The constraints $W^c=0$ can be solved simply by setting
$w_{u\dot\alpha}=\rho\delta_{u\dot\alpha}$, so that
we have
\begin{equation}
\label{delta 2x2}
\Delta(\hat x)=\begin{pmatrix}
\rho\\
{\hat x}-{a'}
\end{pmatrix}
\qquad ,\qquad
\overline\Delta(\hat x)=\begin{pmatrix}
\rho
&,&
\overline{{\hat x}}
-{\bar{a'}}
\end{pmatrix}~~.
\end{equation}
With this choice one can check that Eq. (\ref{ADHM1}) is
satisfied with
\begin{equation}
\frac{1}{f({\hat x})}=\rho^2+|{\hat x}-a'|^2
\end{equation}
where $|{\hat x}-a'|^2=\sum_{\mu}({\hat x}_\mu-a'_\mu)^2$.

To proceed further, we need to find a null vector for $\overline \Delta(\hat x)$.
A solution of $\overline \Delta(\hat x)\,U(\hat x)=0$ has been proposed in
Ref. \cite{Chu:2001cx} where a non-commutative instanton configuration has
been obtained in the regular gauge.
However, to make contact with the string theory
realization of the gauge instantons presented in sections
\ref{sec:d3d-1wb} and \ref{secn:solution}, we need to have the gauge
vector field in the singular gauge since the entire instanton charge is
concentrated at the locations of the D$(-1)$ branes.
Therefore, we now present a different expression for $U(\hat x)$
that we derive by mimicking what is usually done in commutative
ADHM construction to obtain the instanton profile in the singular gauge.
More specifically, we consider the matrix
\begin{equation}
\label{U singular}
U(\hat x)=\begin{pmatrix}
\frac{|{\hat x}-a'|}{\sqrt{\rho^2+|{\hat x}-a'|^2}}\\
-\rho(\hat x-a')\,\frac{1}{|{\hat x}-a'|\sqrt{\rho^2+|{\hat x}-a'|^2}}\\
\end{pmatrix}
\end{equation}
which is the straightforward generalization
of the matrix used in the commutative case, where
the coordinates $x^\mu$ have been replaced by the operators $\hat x^\mu$. With
some simple manipulations one can check that $U(\hat x)$ is a null
vector of $\overline \Delta(\hat x)$. However, it does not
satisfy the completeness constraint (\ref{ADHM2}), as noticed%
\footnote{See however \cite{Tian:2002si}, where a modification of the ansatz is proposed.}
 in \cite{Lechtenfeld:2001ie}, in agreement
with the impossibility of finding a gauge transformation from the
regular to the singular gauge in non-commutative theories
\cite{Chu:2001cx}. Despite this fact, the matrix
(\ref{U singular}) can still be used to obtain an instanton
profile under special conditions.

To see this, it is useful to exploit the one-to-one correspondence
between operators depending on the non-commuting ${\hat x}^\mu$'s and
ordinary functions of $x^\mu$ multiplied with the Moyal product
\begin{equation}
F(x)\star G(x)\equiv F(x)
\exp\left\{\frac{\ii}{2}\,\theta^{\mu\nu}\,{\overleftarrow{\partial_\mu}}\,
{\overrightarrow{\partial_\nu}}\right\}G(x)~~.
\end{equation}
The precise correspondence is obtained by taking the Fourier transform of a
function of $x^\mu$ and anti-transforming it back with ${\rm e}^{-\ii k\cdot{\hat x}}$.
In particular, following this rule one can show that $x^\mu$ corresponds simply
to $\hat{x}^\mu$, and more generally that to any function it corresponds
an operator constructed with the complete symmetrization of the ${\hat
x}^\mu$'s. For example, the function $x^\mu\,x^\nu$ corresponds to
$\frac{1}{2}\left[{\hat x}^\mu{\hat x}^\nu+{\hat x}^\nu{\hat
x}^\mu\right]$, while $|x-a'|^2$ corresponds to $|{\hat
x}-a'|^2$. Along these lines, it is possible to show
that
\begin{equation}
\label{corrcompl}
\frac1{(\rho^2+|{\hat x}-a'|^2)^\alpha}~~
\leftrightarrow~~
2\,\frac{(2|\theta|)^{-\alpha}}{\Gamma(\alpha)}
\int^1_0\hspace{-.2cm}
dt\left(\frac{1-t}{1+t}\right)^{\frac{\rho^2}{2|\theta|}}
\frac{\ee^{-t\frac{|x-a'|^2}{|\theta|}}}{\log^{1-\alpha}\left(\frac{1+t}{1-t}
\right)}
\end{equation}
for any $\alpha>0$, from which (for $\rho=a'=0$ and $\alpha=1$) it follows  that
\begin{equation}
\label{corrxm2}
\frac{1}{|{\hat x}|^2}
\leftrightarrow
\frac{1}{|x|^2}\left(1-\ee^{-\frac{|x|^2}{|\theta|} }\right)~~.
\end{equation}

Let us now apply this correspondence to the matrix $U(\hat x)$ in
\eqref{U singular} and fix, for simplicity, the instanton center in the origin
by setting $a'=0$. Then, one can show that
\begin{equation}
\Delta(x)\star f(x)\star \overline\Delta(x) + U(x)\star\overline
U(x) = \uno + D(x)~~,
\label{violation}
\end{equation}
{\it i.e.} the non-commutative completeness relation \eqref{ADHM2}
is violated by
\begin{equation}
D(x) = \begin{pmatrix}
0~ & ~0\\
0~& ~d(x)
\end{pmatrix}
\qquad{\rm where}\qquad
d(x)=-2\,\ee^{-\frac{|x|^2}{|\theta|}}\begin{pmatrix}
1+\frac{\theta}{|\theta|} & 0\\
0 & 1-\frac{\theta}{|\theta|}
\end{pmatrix}~~.
\label{violation1}
\end{equation}
Notice, however, that this violating term vanishes asymptotically in the large distance
expansion $|x|^2/|\theta|\to \infty$ and is non-perturbative in the non-commutativity parameter.
Thus, if we want to establish a connection with the string results of section
\ref{secn:solution} that have been obtained in the large distance approximation and
with a perturbative expansion in $\theta$, it is natural
to neglect $D(x)$ and use the Moyal counterpart
of \eqref{U singular} to obtain a non-commutative instanton
profile in the singular gauge.
Proceeding in this way, after some tedious algebra, we find
\begin{equation}
\label{finale}
\begin{aligned}
A_\mu(x)=&\left\{\frac{2\rho^2\,\overline\eta^a_{\mu\nu}x^\nu}{|x|^2\rr}\left[1-\frac{
\rho^2(8|x|^4+5|x|^2\rho^4+\rho^4)}
{2|x|^4\rr^3}\,\theta^2\right]\right\}\frac{\tau^a}{2}\\
&+\left\{-\frac{\rho^4\,\theta_{\mu\nu}x^\nu}{|x|^4\rr^2}\right\}\frac{\uno}{2}\,+\,
\mathcal O\left(
\theta^3\right)\\
=&\left\{\frac{2\rho^2}{|x|^4}\, \overline\eta^c_{\mu\nu}x^\nu\left[1-\frac{
\rho^2}{|x|^2}
\left(1+\frac{4\theta^2}{|x|^4}\right)\right]\right\}\frac{\tau^a}{2}\\
&+\,
\left\{-\frac{\rho^4}{|x|^8}\, \theta_{\mu\nu}x^\nu\right\}\frac{\uno}{2} \,+\,
\mathcal O\left(\theta^3
\right)\,+\,
\mathcal O\left(\rho^6
\right)
\end{aligned}
\end{equation}
where in the second step we have supposed
$\frac{\rho^2}{|x|^2}\sim\frac{|\theta|}{|x|^2}\ll 1$.
In our normalization, the quantities in braces represent
the $\mathrm{SU}(2)$ and $\mathrm{U}(1)$ components $A^a_\mu$ and $A^0_\mu$ of the
gauge connection, which
completely agree with the string theory results presented in section \ref{subsec:profile},
and in particular in Eqs. \eqref{gf1}, \eqref{dxres} and
\eqref{udxres2}. Thus, this analysis confirms that the gluon
emission amplitude (\ref{dia1}) from a mixed disk is the correct source for the
non-commutative gauge instantons.

\vskip 0.8cm
\section{Conclusions}
\label{secn:concl}
As we have explained in the previous sections, our
string realization of the non-commutative (anti-)instantons
requires a (anti-)self-dual background $B$ field. In fact, only in
such a background the D3/D$(-1)$ is stable and allows for an exact conformal
field theory description in terms of twisted fields.
Therefore, it is natural to ask what happens in a generic
background and to what extent the D brane realization can be used in this
case. To answer this question one can use
a perturbative approach, similarly to what
has been done for the RR backgrounds giving rise to non-anti-commutative
deformations \cite{Billo:2004zq}. In other words one can treat a generic
$B$ background as a perturbation around flat space and deduce its effects on the
gauge instantons by computing mixed amplitudes with insertions of open string
vertex operators, describing the ADHM moduli, and of the closed string
vertex operator, describing a constant NS-NS $B$ field. Up to a suitable
normalization, the latter is (in the $(-1)$ superghost picture)
\begin{equation}
V_B ~\simeq ~B_{\mu\nu}\,\big(\psi_L^\mu\,{\rm e}^{-\phi_L}\big)\,
\big(\psi^\nu_R
\,{\rm e}^{-\phi_R}\big)
\label{vb}
\end{equation}
where the subscripts $L$ and $R$ denote the left and right moving
parts of the closed string coordinates and superghosts.

The simplest mixed open/closed string diagram
involves one vertex $V_D$ for the auxiliary moduli $D_C^-$ given
in (\ref{vertDmod}) and one closed string vertex $V_B$. Using standard
conformal field theory methods, one finds that the
amplitude under consideration is
\begin{equation}
\begin{aligned}
\lvev V_D V_B \rvev &\equiv
C_0 \int \frac{dy\, d^2z}{dV_{\mathrm{CKG}}}
~
\langle V_D(y)\,V_B(z,\overline z) \rangle
\\
& \simeq~
\frac{\ii}{g_0^2\,(2\pi\alpha')}\,D_{c}^-\,\overline\eta^c_{\mu\nu} B^{\mu\nu}
\end{aligned}
\label{db}
\end{equation}
where we have explicitly exhibited all dimensional constants, but
neglected numerical factors which could be absorbed into the
normalization of the vertex $V_B$.

The amplitude (\ref{db}) turns out to be the only one that is relevant in the
limit $\alpha'\to 0$ with $g_0$ kept fixed, which is the appropriate
field theory limit for disk diagrams involving open strings with at least
one end-point on the
D-instantons \cite{Billo:2002hm}. Indeed, all other amplitudes with different
open string vertex operators or with more insertions of the closed
string vertex $V_B$ either vanish or are sub-leading with respect
to (\ref{db}) in the field theory limit.
Using (\ref{g0gym}) and (\ref{theta}), we can rewrite the above
result as
\begin{equation}
\lvev V_D V_B \rvev~ \simeq~\ii\, D_{c}^-\,\overline\eta^c_{\mu\nu}
\,(2\pi\alpha'B^{\mu\nu}) ~=~
 \ii\,D_{c}^-\,\overline\eta^c_{\mu\nu}
\theta^{\mu\nu}
\label{dtheta}
\end{equation}
which shows that the bosonic ADHM constraint (\ref{bosADHM}) gets
modified by the addition of a term proportional to
$\overline\eta^c_{\mu\nu}\,\theta^{\mu\nu}$.
Of course, if the $B$ background is self-dual this term vanishes in agreement with what we
already found in section \ref{subsec:mod_ADHM}. On the contrary, in an
anti-self-dual background the deformation corresponding to (\ref{dtheta})
is non-trivial and agrees with the explicit ADHM analysis for non-commutative gauge
theories.

In conclusion we have shown that the D3/D$(-1)$ system in a NS-NS
$B$ field correctly describes the instantons of non-commutative
gauge theories. When the background and the gauge field strengths
have the same duality, the brane system is stable and the non-commutative
deformation can be treated exactly by means of a twisted conformal
field theory; in the other cases only a perturbative string approach is
available, but in the field theory limit this always agrees with the non-commutative ADHM
construction. Finally, it would be interesting to explore, both in string and in field theory,
the meaning of the non-perturbative corrections in $|\theta|/x^2$
to the matrix $U(x)$ of the non-commutative ADHM construction that
are needed to satisfy exactly the completeness relation in the singular gauge.

\vskip 1.5cm
\noindent {\large {\bf Acknowledgments}}
\vskip 0.2cm
\noindent We thank Francesco Fucito and Igor Pesando for many
useful discussions.

\vspace{0.8cm}
\appendix
\section{Notation and Conventions}
\label{app:a}

\paragraph{Spinor notation in Euclidean $\mathbb R^4$:}
\label{subapp:spinor}

We consider the Euclidean space $\mathbb R^4$ with coordinates $x^\mu$,
$\mu=1,\cdots 4$. The Clifford algebra $\{\gamma^\mu,\gamma^\nu\} =2\delta^{\mu\nu}$
is satisfied by
\begin{equation}
\gamma^\mu=(\gamma^\mu)^\dag=
\begin{pmatrix}
0 & \sigma^\mu\\
\bar\sigma^\mu & 0
\end{pmatrix}
\end{equation}
where $\sigma^\mu=(\ii\vec\tau,\uno)$,
$\bar\sigma^\mu=(\sigma^\mu)^\dag=(-\ii\vec\tau,\uno)$, with
$\vec\tau$ being the Pauli matrices.

The matrices $\bar\sigma^\mu$ and $\sigma^\mu$, which satisfy the
algebra
\begin{equation}
\label{clifford sigma}
\sigma^\mu\bar\sigma^\nu+\sigma^\nu\bar\sigma^\mu=
\bar\sigma^\mu\sigma^\nu+\bar\sigma^\nu\sigma^\mu=2\delta^{\mu\nu}\,\uno~~,
\end{equation}
are the Weyl matrices acting, respectively, on chiral and anti-chiral
spinors $\psi_{\alpha}$ and $\psi^{\dot\alpha}$.

In our basis, the charge conjugation matrix $C$
is block diagonal and is given by
\begin{equation}
C=-\begin{pmatrix}
\epsilon^{\alpha\beta} & 0\\
0 & \epsilon_{\dot\alpha\dot\beta}
\end{pmatrix}~.
\end{equation}
where
$\epsilon^{12}=\epsilon_{12}=-\epsilon^{\dot 1\dot 2}=
-\epsilon_{\dot 1\dot 2}=1$.
We use the convention that the dotted indices are contracted in the
$\nearrow$ direction, while undotted ones contracted in the $\searrow$ direction:
\begin{equation}
\psi^\alpha=\epsilon^{\alpha \beta}\psi_\beta~~~,~~~
\psi_{\dot\alpha}=\epsilon_{\dot\alpha\dot
\beta}\psi^{\dot\beta}~~.
\end{equation}
Out of the $\gamma$ matrices we may construct the generators
of the $\mathrm{so}(4)$ algebra
\begin{equation}
\Sigma^{\mu\nu}=\frac{1}{4}\left[\gamma^\mu,\gamma^\nu\right]=
\begin{pmatrix}
\frac{1}{2}\sigma^{\mu\nu} & 0\\
0&\frac{1}{2}\bar\sigma^{\mu\nu}
\end{pmatrix}
\end{equation}
where
\begin{equation}
\label{sigmamunu}
\sigma^{\mu\nu}
=\frac{1}{2}\big(\sigma^\mu\bar\sigma^\nu-\sigma^\nu\bar\sigma^\mu\big)
~~~~,~~~~
\bar\sigma^{\mu\nu}
=\frac{1}{2}\big(\bar\sigma^\mu\sigma^\nu-\bar\sigma^\nu\sigma^\mu\big)
\end{equation}
which are respectively self-dual and anti-self-dual.
They act on the irreducible spinor representations corresponding,
respectively, to chiral and anti-chiral spinor, as follows
\begin{equation}
\label{trasfpsi}
\delta_\omega \psi_\alpha = \frac 14 \omega_{\mu\nu}\,
(\sigma^{\mu\nu})_\alpha^{\phantom\alpha\beta}\psi_\beta
~~~~,~~~~
\delta_\omega \psi^{\dot\alpha} = \frac 14 \omega_{\mu\nu}\,
(\overline\sigma^{\mu\nu})^{\dot\alpha}_{\phantom\alpha\dot\beta}
\psi^{\dot\beta
}~~.
\end{equation}
Furthermore, using the symmetry properties of the matrices
$\sigma^{\mu\nu}$ and $\overline \sigma^{\mu\nu}$,
one can check that also $\psi^{\beta}$ and  $\psi_{\dot\beta}$
transform as chiral and anti-chiral spinors:
\begin{equation}
\label{trasfpsi2}
\delta_\omega \psi^\alpha = -\frac 14 \omega_{\mu\nu}\,
\psi^\beta (\sigma^{\mu\nu})_\beta^{\phantom\beta\alpha}
~,\hskip 0.8cm
\delta_\omega \psi_{\dot\alpha} = -\frac 14 \omega_{\mu\nu}\,
\psi_{\dot\beta}(\overline\sigma^{\mu\nu})^{\dot\beta}_
{\phantom\beta\dot\alpha}~.
\end{equation}

The anti-hermiticity of the  $\sigma^{\mu\nu}$ matrices:
${[(\sigma^{\mu\nu})^{\phantom\alpha\beta}_\alpha]}^*$
$=-(\sigma^{\mu\nu})^{
\phantom\alpha\alpha}_\beta$
implies instead that $(\psi_\alpha)^*$ transform like
$\psi^\alpha$.
Then, consistently  with the conjugation property of
$\sigma^\mu$, we define the complex conjugation as
\begin{equation}
\label{conjugation}
\begin{aligned}
(\psi^\alpha)^*&=\overline\psi_\alpha\qquad,\qquad
&&(\psi_{\dot\alpha})^*=-\overline\psi^{\dot\alpha}\\
(\psi_\alpha)^*&=-\overline\psi^\alpha\qquad,\qquad
&&(\psi^{\dot\alpha})^*=\overline\psi_{\dot\alpha}~~.\\
\end{aligned}
\end{equation}

\paragraph{'t Hooft symbols:}
\label{subapp:thooft}
The $\mathrm{Spin}(4)$ group is isomorphic to $\mathrm{SU}(2)_+\times
\mathrm{SU}(2)_-$. The isomorphism is described at the level of
generators using the 't Hooft symbols:
\begin{equation}
\label{isomorphism}
J_{\mu\nu} = \eta^c_{\mu\nu} J^+_c + \overline \eta^c_{\mu\nu} J^-_c~,
\end{equation}
where $J_{\mu\nu}$ ($\mu,\nu=1,\ldots,4$) are the $\mathrm{Spin}(4)$
generators and $J^\pm_c$ ($c=1,2,3$) are the $\mathrm{SU}(2)_\pm$
generators.
Explicitly, the 't Hooft symbols are defined as follows:
\begin{subequations}
\begin{align}
&\overline\eta^c_{\mu\nu}=-\bar\eta^c_{\nu\mu}~,\qquad\eta^c_{\mu\nu}
=-\eta^c_ { \nu\mu}~,\\
&\eta^c_{ab}=\bar\eta^c_{ab}=\epsilon_{cab}~,\qquad a,b,c\in\{1,2,3\}~,\\
&\overline\eta^c_{4a}=\eta^c_{a4}=\delta^c_a
\end{align}
\end{subequations}
so that $\eta $ is self-dual and $\bar\eta$ is anti-self-dual.

Applied to the irreducible spinor representations, \eqref{isomorphism}
states that chiral and anti-chiral spinors belong respectively to the
fundamental representation of $\mathrm{SU}(2)_+$ and $\mathrm{SU}(2)_-$:
\begin{equation}
\label{iso_fun}
(\sigma_{\mu\nu})^{\phantom\alpha\beta}_\alpha=
\ii\eta^c_{\mu\nu}(\tau^c)^{\phantom\alpha\beta}_\alpha\qquad,\qquad
(\bar\sigma_{\mu\nu})^{\dot\alpha}_{\phantom\alpha\dot\beta}=
\ii\bar\eta^c_{\mu\nu}(\tau^c)^{\dot\alpha}_{\phantom\alpha\dot\beta}~.
\end{equation}

Let us collect here some useful formulae for manipulating the 't Hooft
symbols:
\begin{eqnarray}
\label{contetamunu}
&\eta^c_{\mu\nu}\eta^d_{\mu\nu}=\bar\eta^c_{\mu\nu}\bar\eta^d_{\mu\nu}
=4\delta^{cd}\\
\label{deltaeta}
&\eta^c_{\mu\nu}\eta^c_{\rho\sigma}=
\delta_{\mu\rho}\delta_{\nu\sigma}-\delta_{\nu\rho}\delta_{\mu\sigma}
+\epsilon_{\mu\nu\rho\sigma}
\\
&\bar\eta^c_{\mu\nu}\bar\eta^c_{\rho\sigma}=
\delta_{\mu\rho}\delta_{\nu\sigma}-\delta_{\nu\rho}\delta_{\mu\sigma}
-\epsilon_{\mu\nu\rho\sigma}
\end{eqnarray}
We also have
\begin{equation}
\label{epsietab}
\epsilon_{abc} \overline\eta^{b}_{\nu\sigma} \overline\eta^{c}_{\rho\tau} =
\overline\eta^{a}_{\sigma\tau} \delta_{\nu\rho} +
\overline\eta^{a}_{\nu\rho} \delta_{\sigma\tau} -
\overline\eta^{a}_{\nu\tau} \delta_{\sigma\rho} -
\overline\eta^{a}_{\sigma\rho} \delta_{\nu\tau}~,
\end{equation}
and similarly for the $\eta^{b}_{\nu\sigma}$ symbols.



\begin{thebibliography}{99}

\bibitem{Connes:1997cr}
  A.~Connes, M.~R.~Douglas and A.~Schwarz,
  JHEP {\bf 9802} (1998) 003
  [arXiv:hep-th/9711162].

\bibitem{Abouelsaood:1986gd}
A. Abouelsaood, C. G. Callan, C. R. Nappi and S. A. Yost,
Nucl.\ Phys.\ B {\bf 280} (1987) 599.

\bibitem{Sheikh-Jabbari:1997yi}
  M.~M.~Sheikh-Jabbari,
  Phys.\ Lett.\ B {\bf 425} (1998) 48
  [arXiv:hep-th/9712199].

\bibitem{Chu:1998qz}
  C.~S.~Chu and P.~M.~Ho,
  Nucl.\ Phys.\ B {\bf 550} (1999) 151
  [arXiv:hep-th/9812219].

\bibitem{Schomerus:1999ug}
  V.~Schomerus,
  JHEP {\bf 9906} (1999) 030
  [arXiv:hep-th/9903205].

\bibitem{Ardalan:1999av}
F. Ardalan, H. Arfaei, H. and M. M. Sheikh-Jabbari,
Nucl.\ Phys.\ B{\bf 576} (2000) 578
[arXiv:hep-th/9906161]

\bibitem{Chu:1999gi}
  C.~S.~Chu and P.~M.~Ho,
  Nucl.\ Phys.\ B {\bf 568} (2000) 447
  [arXiv:hep-th/9906192].


\bibitem{Seiberg:1999vs}
  N.~Seiberg and E.~Witten,
  JHEP {\bf 9909} (1999) 032
  [arXiv:hep-th/9908142].

\bibitem{Chu:1999ta}
  C.~S.~Chu,
  ``Noncommutative open string: Neutral and charged''
  [arXiv:hep-th/0001144].

\bibitem{Haggi-Mani:2000uc}
P.~Haggi-Mani, U.~Lindstrom and M.~Zabzine,
Phys.\ Lett.\ B {\bf 483} (2000) 443
[arXiv:hep-th/0004061].

\bibitem{Mihailescu:2000dn}
  M.~Mihailescu, I.~Y.~Park and T.~A.~Tran,
  Phys.\ Rev.\ D {\bf 64} (2001) 046006
  [arXiv:hep-th/0011079].

\bibitem{Chu:2000wp}
  C.~S.~Chu, R.~Russo and S.~Sciuto
  Nuc. Phys. B {\bf 585} (2000) 193
  [arXiv:hep-th/0004183]

\bibitem{Chu:2000pc}
C.~S.~Chu and R.~Russo,
Mod.\ Phys.\ Lett.\ A {\bf 16} (2001) 211
[Fortsch.\ Phys.\  {\bf 49} (2001) 633]
[arXiv:hep-th/0102164].

\bibitem{deBoer:2003dn}
J.~de Boer, P.~A.~Grassi and P.~van Nieuwenhuizen,
Phys.\ Lett.\ B {\bf 574} (2003) 98
[arXiv:hep-th/0302078].

\bibitem{Ooguri:2003qp}
H.~Ooguri and C.~Vafa,
Adv.\ Theor.\ Math.\ Phys.\  {\bf 7} (2003) 53
[arXiv:hep-th/0302109];
  Adv.\ Theor.\ Math.\ Phys.\  {\bf 7} (2004) 405
  [arXiv:hep-th/0303063].

\bibitem{Seiberg:2003yz}
N.~Seiberg,
JHEP {\bf 0306} (2003) 010
[arXiv:hep-th/0305248].

\bibitem{Witten:1995im}
E.~Witten,
Nucl.\ Phys.\ B {\bf 460} (1996) 335
[arXiv:hep-th/9510135].

\bibitem{Douglas}
M.~R.~Douglas,
J.\ Geom.\ Phys.\  {\bf 28}, 255 (1998)
[arXiv:hep-th/9604198];
``Branes within branes''
[arXiv:hep-th/9512077].

\bibitem{Atiyah:ri}
  M.~F.~Atiyah, N.~J.~Hitchin, V.~G.~Drinfeld and Y.~I.~Manin,
  Phys. Lett. A {\bf 65} (1978) 185.

\bibitem{Dorey:2002ik}
  N.~Dorey, T.~J.~Hollowood, V.~V.~Khoze and M.~P.~Mattis,
  Phys. Rept. {\bf 371} (2002) 231
  [arXiv:hep-th/0206063].

\bibitem{Billo:2002hm}
  M.~Bill\`o, M.~Frau, I.~Pesando, F.~Fucito, A.~Lerda and A.~Liccardo,
  JHEP {\bf 0302} (2003) 045
  [arXiv:hep-th/0211250].

\bibitem{Green:2000ke}
  M.~B.~Green and M.~Gutperle,
  JHEP {\bf 0002} (2000) 014
  [arXiv:hep-th/0002011].

\bibitem{Witten:1995gx}
  E.~Witten,
  Nucl.\ Phys.\ B {\bf 460}, 541 (1996)
  [arXiv:hep-th/9511030].

\bibitem{Nekrasov:1998ss}
  N.~Nekrasov and A.~Schwarz
  Commun. Math. Phys {\bf 198} (1998) 689
  [arXiv:hep-th/9802068].

\bibitem{Furuuchi1}
K.~Furuuchi,
Prog.\ Theor.\ Phys.\  {\bf 103} (2000) 1043
[arXiv:hep-th/9912047].

\bibitem{Furuuchi2}
K.~Furuuchi,
``Topological charge of U(1) instantons on noncommutative $R^4$'',
[arXiv:hep-th/0010006].

\bibitem{Nekrasov:2000zz}
N. A. Nekrasov,
Commun. Math. Phys., {\bf 241} (2003) 143-160
 [arXiv:hep-th/0010017].

\bibitem{Furuuchi3}
K.~Furuuchi,
JHEP {\bf 0103} (2001) 033
[arXiv:hep-th/0010119].

\bibitem{ND}
M.~R.~Douglas and N.~A.~Nekrasov,
``Noncommutative field theory''
[arXiv:hep-th/0106048].

\bibitem{Chu:2001cx}
  C.~S.~Chu, V.~V.~Khoze and G.~Travaglini,
  Nucl.\ Phys.\ B {\bf 621} (2002) 101
  [arXiv:hep-th/0108007].

\bibitem{Lechtenfeld:2001ie}
  O.~Lechtenfeld and A.~D.~Popov,
  JHEP {\bf 0203} (2002) 040
  [arXiv:hep-th/0109209].

\bibitem{Tian:2002si}
  Y.~Tian and C.~J.~Zhu,
  Phys.\ Rev.\ D {\bf 67} (2003) 045016
  [arXiv:hep-th/0210163].

\bibitem{Wimmer:2005bz}
  R.~Wimmer,
  JHEP {\bf 0505} (2005) 022
  [arXiv:hep-th/0502158].

\bibitem{Billo:2004zq}
  M.~Bill\`o, M.~Frau, I.~Pesando and A.~Lerda
  JHEP {\bf 0405} (2004) 023
  [arXiv:hep-th/0402160].

\bibitem{Billo:2005jw}
  M.~Bill\`o, M.~Frau, F.~Lonegro and A.~Lerda
  JHEP {\bf 0505} (2005) 047
  [arXiv:hep-th/0502084].


\bibitem{Dixon:jw}
  L.~J.~Dixon, J.~A.~Harvey, C.~Vafa and E.~Witten,
  Nucl.\ Phys.\ B {\bf 261} (1985) 678;
  S.~Hamidi and C.~Vafa,
  Nucl.\ Phys.\ B {\bf 279} (1987) 465.

\bibitem{Belavin:fg}
  A.~A.~Belavin, A.~M.~Polyakov, A.~S.~Schwarz and Y.~S.~Tyupkin,
  Phys. Lett. B {\bf 59}, 85 (1975);
  G.~'t Hooft,
  Phys. Rev. D {\bf 14}, 3432 (1976)
  [Erratum-ibid. D {\bf 18}, 2199 (1978)].

\bibitem{Armoni:2000xr}
  A.~Armoni,
  Nucl.\ Phys.\ B {\bf 593} (2001) 229
  [arXiv:hep-th/0005208].

\bibitem{Bonora:2000ga}
  L.~Bonora and M.~Salizzoni,
  Phys.\ Lett.\ B {\bf 504} (2001) 80
  [arXiv:hep-th/0011088].



\end{thebibliography}
\end{document}